\documentclass{elsart}
\usepackage{graphicx}

\def\grs{$\gamma$-rays \,}
\def\gr{$\gamma$-ray \,}

\begin{document}
 
\runauthor{}
\begin{frontmatter}
\title{Similarities and Differences  between  Relativistic Electron-Photon 
Cascades  Developed in Matter, Photon Gas and Magnetic Field}

\author[mpik]{F.A. Aharonian\thanksref{1}}
\author[altai]{A.V. Plyasheshnikov\thanksref{2}}

\thanks[1]{E-mail: Felix.Aharonian@mpi-hd.mpg.de}
\thanks[2]{E-mail: plya@theory.dcn-asu.ru}
 
\address[mpik]{MPI f\"ur Kernphysik, Sauperfercheckweg 1,
D-69117 Heidelberg, Germany}
\address[altai]{Altai State University, Dimitrov street 66, Barnaul, 
656099, Russia}

\begin{abstract}                                                               
We  investigate the properties of astrophysical electromagnetic  cascades 
in matter,  photon gas   and  magnetic fields, and  discuss  similarities 
and differences between characteristics of electron-photon showers 
developed in these 3 substances.  We apply  the same computational 
technique based on  solution of the  adjoint cascade equations to all 
3 types of cascades, and  present precise numerical 
calculations of  cascade curves  and broad-band energy spectra of 
secondary electrons  and  photons  at different penetration lengths.  
\end{abstract}   
 
\begin{keyword}
Electromagnetic Cascades, High Energy Processes, Gamma-Ray Sources
\end{keyword}
\end{frontmatter}                                                              

\section{Introduction}

Relativistic electrons --  directly accelerated,  or being secondary 
products of various hadronic processes -- may result in copious  
$\gamma$-ray production  caused by  interactions with ambient 
targets in forms of {\em gas (plasma),  radiation} and {\em magnetic fields}.
In  different astrophysical environments $\gamma$-ray 
production may proceed  with high efficiency 
through {\em bremsstrahlung},  {\em inverse Compton scattering} and  
{\em synchrotron (and/or  curvature) radiation},   respectively.  
Generally,  $\gamma$-ray  production is  effective  when 
the cooling time that characterizes   the  rate of the process 
does  not  significantly  exceed {\em (i)} the  source (accelerator) age 
 {\em (ii)} the  characteristic time of  non-radiative 
losses  caused by   adiabatic expansion of the source or  particle 
escape  and   {\em (iii)} the cooling  time of  competing  radiation mechanisms  that 
result in  low-energy photons  {\em outside}  the $\gamma$-ray domain.
As long as the charged particles are effectively confined  
in the $\gamma$-ray production region, at some circumstances 
these  condition   could be  fulfilled  even in environments  with  
a relatively low  gas and photon densities  or weak magnetic field.
More specifically,   the $\gamma$-ray production efficiency could be 
close to 1 even when  $t_{\rm rad} \gg R/c$ ($R$ is  the characteristic 
linear size of the production region, $c$ is the speed of light).
In  such  cases the secondary  $\gamma$-rays  escape 
the  source  without significant internal absorption.
Each of the above mentioned $\gamma$-ray {\em production} mechanisms  
has its major ``counterpart'' - $\gamma$-ray {\em absorption} mechanism  
of same electromagnetic origin resulting in  electron-positron pair production 
in matter (the counterpart of bremsstrahlung), in photon gas 
(the counterpart of inverse Compton scattering), and in
magnetic field (the counterpart of synchrotron radiation). 
The  $\gamma$-ray production  mechanisms and 
their absorption counterparts have similar cross-sections, therefore 
the condition  for radiation $t_{\rm rad}  \geq  R/c$ generally
implies small optical depth for the corresponding $\gamma$-ray 
absorption mechanism, $\tau_{\rm abs} \leq  1$.  
But in many astrophysical  scenarios, in particular in compact  
galactic and extragalactic objects  with  favorable conditions 
for particle acceleration,  the radiation processes 
proceed so fast that $t_{\rm rad}  \geq  R/c$. At these conditions 
the internal $\gamma$-ray absorption  becomes unavoidable.   
If  the $\gamma$-ray production and absorption  processes 
occur  in  relativistic  regime,  namely   when 
(i) $E_{\rm \gamma, e} \geq  10^3  m_e c^2$ in the hydrogen gas, 
(ii) $E_{\rm \gamma,e} \epsilon \gg m_{\rm e}^2c^4$  
in photon gas  (often called Klein-Nishina regime; 
$\epsilon$ is the average energy of the target photons), or 
(iii) $(E_{\rm \gamma,e} /m_{\rm e}c^2)(H/H_{\rm cr}) \gg 1$ 
in the magnetic field (often called quantum regime; 
$H_{\rm cr}\simeq 4.4 \times 10^{13} \ \rm G$ is the so-called critical 
strength of the  magnetic field), the problem cannot be reduced to 
a simple  absorption effect. In this regime,  the secondary electrons 
produce new generation of high energy $\gamma$-rays, these  photons 
again produce electron-positron pairs, so the electromagnetic cascade 
develops.

The  characteristics of electromagnetic cascades  in matter
have been comprehensively studied,  basically  in the context 
of interactions of cosmic rays with the Earth's atmosphere
(see e.g.  \cite{matter})  as well as  for calculations of performance 
of detectors of high energy  particles   (e.g. \cite{EGS4}).  The 
theory of electromagnetic  cascades in matter can be applied to 
some   sources  of high energy cosmic radiation, in particular to 
the  ``hidden source''   scenarios like massive black holes 
in centers of AGN or young pulsars  inside the dense shells of 
recent supernovae explosions  (see e.g. \cite{Berez_book}). 
Within another,   so-called ``beam dump'' models
(see e.g. Ref. \cite{Berez_book,Halzen})
applied to X-ray binaries,  protons accelerated by the compact 
object (a neutron star or black hole), hit 
the atmosphere of the normal 
companion star, and thus  result in   production of high energy 
neutrinos and  $\gamma$-rays \cite{Ber_Vest}. 
In such objects,  the thickness of the surrounding  gas 
can significantly exceed  $100 \ \rm g/cm^2$, thus 
the  protons    produced  in the central source would  initiate
(through production of high energy $\gamma$-rays and electrons) 
electromagnetic showers.   
These  sources  perhaps represent the ``best hope'' 
of neutrino astronomy,  but they 
are generally considered as less  attractive  targets for 
gamma-ray astronomy.  However,  the $\gamma$-ray 
emission  in  these objects is not  fully suppressed. The 
recycled radiation  with spectral features  determined 
by  the  thickness (``grammage'')  of the gas shell, should be seen 
in $\gamma$-rays in any case, unless the synchrotron radiation 
of secondary electrons dominates over the bremsstrahlung losses
and  channels the main  fraction of the nonthermal energy into 
the sub gamma-ray domains.       

The development of electromagnetic cascades in photon gas and 
magnetic fields is  a  more common phenomenon in astrophysics.  
In photon fields  such cascades can be created on  almost {\em all} 
astronomical scales,  from  compact objects like  
accreting black holes \cite{AVK,Zdz,Svenson,CopBl,mastiprot}, 
fireballs in gamma-ray  bursts \cite{GRBs} and   
sub-pc jets of  blazars \cite{blazars} to large-scale (up to $\geq 100$ kpc)
AGN  jets  \cite{bierman,gamma_jets} and  $\geq 1$ Mpc size
clusters of galaxies \cite{psynch}. 
Electromagnetic cascades in the intergalactic medium lead 
to formation  of huge ($\geq 10$ Mpc) nonthermal structures like  
hypothetical electron-positron pair halos  \cite{halos}.
Finally, there is little doubt that the entire Universe is a
scene of  continuous  creation and 
development of  electromagnetic cascades.  
All  \grs of energy $\geq$ a few GeV emitted
by  astrophysical objects   have a similar  fate. Sooner or later 
they terminate  on Hubble scales due to interactions with the 
diffuse extragalactic background. Since 
the energy density of 2.7 K CMBR significantly exceeds the 
 energy density of intergalactic magnetic fields, these interactions 
initiate electron-photon cascades  \cite{intergalactic}.   
The superposition of 
contributions of \grs from these cascades should constitute a significant  
fraction of the observed diffuse extragalactic background.

Bonometto and  Rees \cite{Bo_Rees} perhaps where the first who realized 
the astrophysical importance  of  development of  
electron-photon cascades  supported by $\gamma$-$\gamma$ 
pair-production and  inverse Compton scattering in dense photon fields. 
When the so called compactness parameter  \cite{Guilbert}
$l=(L/R) (\sigma_{\rm T}/m_{\rm e}c^3)$  (L is the luminosity and R 
is the radius  of the source) is less than 10, 
then the cascade is developed  in the 
linear regime,  i.e. when  the soft radiation 
produced by cascade electrons do not have a significant feedback effect
on the cascade development. 
In many cases, including the cascade development in compact objects, 
this approximation works quite well. 

The first quantitative study of characteristic of 
linear cascades in photon fields has been performed  
using  the method of Monte Carlo simulations  \cite{Ah_vard_kir}.
Generally,  the kinetic equations for the cascade particles 
can  be  solved only numerically. However, with some simplifications  
it is possible to derive useful analytical approximations \cite{Zdz,Svenson}
which help to understand the features of  the  steady-state solutions
for  cascades in photon fields. 

The cascade development in the magnetic field is a key element to 
understand the physics of pulsar magnetospheres \cite{sturrock,HardBaring},
therefore it is  generally treated as a process associated with very   strong 
magnetic fields. However, such cascades  could be triggered in many  
other (at first glance unusual) 
sites like the Earth's geomagnetic field \cite{gonchar,Anguelov,Plyah},
accretion disks of massive black holes \cite{Bednarek}, etc.   In general, 
the  pair cascades in  magnetic fields  are  effective 
when  the product of the particle (photon  or electron) energy and the strength
of the field becomes close  to  the ``quantum threshold'' of about 
$H_{\rm crit} m_{\rm e}c^2 \simeq 2 \times 10^7 \ \rm TeV \cdot G$,
unless we assume a  specific,  regular field configuration. 
An approximate approach, similar to the so-called approximation A 
of cascade development in  matter \cite{matter}, has been recently 
applied  by Akhiezer et al. \cite{Akhiezer}. Although this theory quite 
satisfactorily describes  the basic features of photon-electron 
showers, it does not provide  an adequate accuracy 
for quantitative  description of the cascade characteristics \cite{Anguelov}.   
Note that in both studies approximate  cross-sections for magnetic 
bremsstrahlung and magnetic pair-production have  been used, thus  
reducing the validity  of the results  to the limit  
$E H \gg 10^7  \ \rm TeV \cdot G$ ($E$ is the minimum energy of 
secondary particles being under consideration). 

As long as we are interested in the one-dimensional 
cascade development (which seems to be quite  
sufficient for many 
astrophysical purposes),  all 3 types of cascades can be described by 
the same  integro-differential equations like the ones derived by 
Landau and Rumer \cite{Landau},  but  in each case specifying   the 
cross-sections of relevant  interaction processes.  
The solution of these equations in a broad range of energies is 
however not a trivial  task. In this paper we present the results of 
our recent study of cascade characteristics in 3 substances - 
matter, photon gas and magnetic field - 
with emphasis on the analysis of similarities and differences 
between these 3 types of cascades. 
For quantitative studies of these characteristics we have 
chosen the so-called   technique of adjoint cascade equations.     
Although  this  work has been initially motivated by 
methodological and pedagogical  objectives, 
some results  are rather 
original and  may present 
practical interest in certain areas of high energy astrophysics.

\section{Technique of adjoint cascade equations}
The results of this study  are based  on numerical 
solutions  of the  so-called  adjoint cascade 
equations. The  potential of this computational 
technique, in particular its possible applications to
different problems of  cosmic ray physics 
has been  comprehensively described  in 
Ref.\cite{Uchaikin}.  Below we briefly discuss 
the main features of  solution of the  adjoint cascade equations.
For our purposes it is quite sufficient to consider only  the longitudinal 
development of electromagnetic cascades neglecting the emission
angles of secondary particles  and, thus,  assuming  that all cascade 
particles are moved along the shower  axis. Besides,  we  neglect 
the differences  in interactions of electrons and  positrons, i.e. 
do not consider positron  annihilation, both in flight and after 
thermalization  in the ambient plasma. This is a quite good approximation 
as long as we are interested in $\gamma$-ray energies exceeding
$\alpha_{\rm f}^{-1} m_{\rm e}c^2 \simeq 70  \ \rm MeV$
($\alpha_{\rm f}=1/137$ is the fine structure constant).
Under  these assumptions, the system of adjoint 
cascade equations reads
\begin{equation}
\label{eq1}
\frac{\partial f}{\partial t} + f/\lambda_e - 
\int_{\varepsilon_{th}}^{\varepsilon} 
W_{ee}(\varepsilon,\varepsilon') f(t,\varepsilon')d\varepsilon' -
\int_{\varepsilon_{th}}^{\varepsilon} 
W_{e\gamma}(\varepsilon,\varepsilon')g(t,\varepsilon')d\varepsilon' 
= F(t,\varepsilon),
\end{equation}

\begin{equation}
\label{eq2}
\frac{\partial g}{\partial t} + g/\lambda_{\gamma} - 
\int_{\varepsilon_{th}}^{\varepsilon} 
W_{\gamma e}(\varepsilon,\varepsilon') f(t,\varepsilon')d\varepsilon' -
\int_{\varepsilon_{th}}^{\varepsilon} 
W_{\gamma\gamma}(\varepsilon,\varepsilon')g(t,\varepsilon')d\varepsilon' 
= G(t,\varepsilon).
\end{equation}

The adjoint functions $f(t,\varepsilon)$ 
and $g(t,\varepsilon)$ in  Eqs. (\ref{eq1}) 
and (\ref{eq2}) describe  the contributions  (averaged over random 
realizations of the  cascade development)  from  cascade initiated  
by a  primary  electron ($f$) or a photon ($g$) 
of energy $\varepsilon$ (all energies are expressed in units $m_e c^2$); 
$t$ is the cascade penetration depth;
$\varepsilon_{th}$ is the minimum energy of cascade particles being under 
consideration. 

The parameter $\lambda_{\alpha}$  in Eqs. (\ref{eq1}) and  (\ref{eq2})
defines the mean free path length   of cascade particles  of type  $\alpha$
($\alpha=e \ \rm or \ \gamma$). It  can be 
expressed  through the total cross-sections ($\sigma_{\alpha}$) 
of interaction  with the ambient medium. In the most 
general case, when the medium consists of    
matter ($M$),  magnetic field  ($F$) and photon  gas  ($G$),  
%------------------------------
\begin{equation}
\label{eq3}
\lambda_{\alpha}=\left[ n_0^{(M)} \sigma_{\alpha}^{(M)}+
\sigma_{\alpha}^{(F)}+ n_0^{(G)}\sigma_{\alpha}^{(G)}\right]^{-1}
\end{equation}
where $n_0^{(M)}$ and $n_0^{(F)}$  are the number density of  
the matter atoms and the ambient photons, respectively. The 
parameters  
%--------------------------------------------
\begin{equation}
\label{eq4}
W_{\alpha\beta}(\varepsilon,\varepsilon')=
n_0^{(M)} w_{\alpha\beta}^{(M)}(\varepsilon,\varepsilon')+
w_{\alpha\beta}^{(F)}(\varepsilon,\varepsilon') +
n_0^{(G)} w_{\alpha\beta}^{(G)}(\varepsilon,\varepsilon')
\end{equation}
are  differential cross-sections over the energy $\varepsilon'$ of the 
secondary particle of type $\beta$ ($\beta=e$ or $\gamma$). 
They are normalized as
%----------------------------------------------------------------------
\begin{equation}
\label{eq5}
\int W_{\alpha\beta}(\varepsilon,\varepsilon') d\varepsilon' =  
n_0^{(M)} \bar{n}_{\alpha\beta}^{(M)}
\sigma_{\alpha}^{(M)} + 
\bar{n}_{\alpha\beta}^{(F)}
\sigma_{\alpha}^{(F)}  
+ n_0^{(G)} \bar{n}_{\alpha\beta}^{(G)}
\sigma_{\alpha}^{(G)}
\end{equation}
where $\bar{n}_{\alpha\beta}^{(M)}$, $\bar{n}_{\alpha\beta}^{(F)}$  
and $\bar{n}_{\alpha\beta}^{(G)}$    
are the mean multiplicities  
of  secondary particles of type $\beta$  produced  by a  
particle of type $\alpha$.

Properties of  the particle ``detector'' located at the depth 
$t$  are defined by 
the right  hand side functions  $F$ and $G$ in   Eqs.~ (\ref{eq1}) 
and (\ref{eq2}) and by the 
boundary conditions    $f(t=0,\varepsilon)$ and 
$g(t=0,\varepsilon)$. For example, 
if the ``detector'' 
measures the number  of cascade  electrons above some threshold 
 energy,   $\varepsilon \ge \varepsilon_{th}$, then
\begin{equation}
\label{eq41}
F(t,\varepsilon)=G(t,\varepsilon)=0; \quad  
f(t=0,\varepsilon)=H(\varepsilon-\varepsilon_{th}), 
\quad g(t=0,\varepsilon)=0
\end{equation}   
where   
$H(x)=1$ for $\quad  x\ge 0$  and  
$H(x)= 0$ for $\quad x<0$. 
Analogously, if the ``detector'' measures the number  of cascade  
photons with energy  
$\varepsilon \ge \varepsilon_{th}$, then
\begin{equation}
\label{eq42}
F(t,\varepsilon)=G(t,\varepsilon)=0; \quad
\quad f(t=0,\varepsilon)=0, \quad  
g(t=0,\varepsilon)=H(\varepsilon-\varepsilon_{th}). 
\end{equation}

For solution of the adjoint cascade equations we use the  numerical
method  proposed  in Ref. \cite{Plyasheshnikov}. To apply this 
approach to Eqs. (\ref{eq1}) and (\ref{eq2}) we introduce an increasing 
subsequence   $\{\varepsilon_k\}=\varepsilon_0=\varepsilon_{th},
\varepsilon_1,\ldots,\varepsilon_k,\ldots$
of energy points,  and corresponding values 
of adjoint  functions  $\{f_k(t),g_k(t)\}$. Let's write  
now the Lagrange polynomial interpolation formulae for an 
approximate  presentation of functions  $f$ and $g$ inside the energy 
intervals $\Delta\varepsilon_k=(\varepsilon_{k-1},\varepsilon_k)$
for $k=1,2,\ldots$ through their values at the support points  
$\{\varepsilon_k\}$,   
$$
f(\varepsilon,t) \simeq \tilde{f}(\varepsilon,t)=
\sum_{j=k-n}^{k}f_{j}(t) L_{kj}^{n}(\varepsilon), \quad
g(\varepsilon,t) \simeq \tilde{g}(\varepsilon,t)=
\sum_{j=k-n}^{k}g_{j}(t) L_{kj}^{n}(\varepsilon); $$
\begin{equation} 
L_{kj}^n(\varepsilon)=\prod_{r=0}^{n}
(\varepsilon-\varepsilon_{k-r})
(\varepsilon_j-\varepsilon_{k-r})^{-1}, \quad r \ne k-j,
\quad \varepsilon \in \Delta \varepsilon_k
\label{app1}
\end{equation}
where $n\equiv n(k)={\rm min}(N,k)$ and $N$ is the maximum 
available power of the Lagrange  polynomial.

For  the first $k$ energy  intervals $\Delta\varepsilon_k$
the support values from   $f_0,g_0$ to $f_k,g_k$ are 
used; in this case the polynomial power $n$ is equal 
to $k$. For intervals with $k\ge N$     
the support  values 
$f_{k-N},g_{k-N},f_{k-N+1},g_{k-N+1},\ldots$, $f_k,g_k$ are involved 
in the interpolation with use of the power $n=N$. 

Let's  adopt  now $\varepsilon=\varepsilon_k$ in Eqs.(\ref{eq1}) and
(\ref{eq2}), and present the integral members of these equations
in the form of the sum over the energy intervals 
$\Delta\varepsilon_i$  ($i=1,2,\ldots,k$):
\begin{equation}
\int_{\varepsilon_0}^{\varepsilon_k}
W_{\alpha\beta}(\varepsilon_k,\varepsilon)\cdots d\varepsilon =
\sum_{i=1}^{k}\int_{\varepsilon_{i-1}}^{\varepsilon_i}
W_{\alpha\beta}(\varepsilon_k,\varepsilon)\cdots d\varepsilon,
\quad \alpha,\beta=e,\gamma \ .
\end{equation} 
 
Then, after some simple calculations,  we find  the 
following equations
\begin{equation}
\label{app2}
\frac{\partial f_k}{\partial t} + A_k f_k - B_k g_k =F'_k, \quad
\frac{\partial g_k}{\partial t} - C_k f_k +D_k g_k =G'_k
\end{equation}
where
\begin{equation}
A_k=1/\lambda_{e}(\varepsilon_k) - a_{kk}, \quad B_k=b_{kk}, 
\quad C_k=c_{kk},  \quad D_k=1/\lambda_{\gamma}(\varepsilon_k) - d_{kk}, 
\end{equation}
\begin{equation}
F'_k=F_k+\sum_{j=0}^{k-1}[a_{jk} f_j+b_{jk} g_j], \quad 
G'_k=G_k+\sum_{j=0}^{k-1}[c_{jk} f_j+d_{jk} g_j].
\end{equation}
The coefficients $a_{jk},b_{jk},c_{ik}$ and $d_{jk}$ can be 
expressed through  cross-sections of relevant processes:  
$$a_{jk}=\sum_{i=1}^{k}\int_{\varepsilon_{i-1}}^
{\varepsilon_i} W_{ee}(\varepsilon_k,\varepsilon)
L_{ij}^n(\varepsilon) \sum_{s=i-n}^i\delta_{js} d\varepsilon \ , 
$$
\begin{equation}   
b_{jk}=\sum_{i=1}^{k}\int_{\varepsilon_{i-1}}^
{\varepsilon_i} W_{e\gamma}(\varepsilon_k,\varepsilon)
L_{ij}^n(\varepsilon) \sum_{s=i-n}^i\delta_{js}d\varepsilon \ , \  etc.
\end{equation}
with $\quad \delta_{js}=1 \,\, {\rm if} \,\, j=s,$ and
$\delta_{js}=0 \,\, {\rm if} \,\, j\ne s$.

Thus, this  method allows us to reduce the  
integro-differential Eqs.  (\ref{eq1})
and (\ref{eq2}) to the system of ordinary differential 
equations (\ref{app2}). 
The solution of   Eq. (\ref{app2}) can be obtained in two steps:

(1)  for  $\varepsilon=\varepsilon_0=\varepsilon_{th}$ in
formulae (\ref{eq1}) and (\ref{eq2}),  we  find  the 
following  equations for functions
$f_0(t)$ and $g_0(t)$
\begin{equation}
\label{app5}
\frac{\partial f_0}{\partial t} + f_0(t)/\lambda_{e}
(\varepsilon_0) = F(t,\varepsilon_0), \quad
\frac{\partial g_0}{\partial t} + g_0(t)/\lambda_{\gamma}
(\varepsilon_0)= G(t,\varepsilon_0)
\end{equation}

(2)  after solving  Eqs. (\ref{app5}) we  can 
calculate the functions 
$f_1(t),g_1(t)$, because for $k=1$  Eq. (\ref{app2}) contains 
only $f_0(t),g_0(t)$, $f_1(t)$ and $g_1(t)$;  
after that one can find by the same way the functions
$f_2(t),g_2(t)$, $f_3(t),g_3(t)$, etc.      

For a fixed value of $k$,  we introduce an increasing  subsequence 
of the depth values $\{ t_l\}$  ($t_0=0, t_l=t_{l-1}+\tau_l$) and 
correspondingly  $f_k(t_l)=f_{k,l}$, $g_k(t_l)=g_{k,l}$.
For each interval  $(t_{l-1}, t_l)$  one can 
evaluate  $f_{k,l},g_{k,l}$ by  solving   Eq. (\ref{app2}) 
for which the values $f_{k,l-1}, g_{k,l-1}$ serve as 
boundary conditions. This leads to 
$$ f_{k,l}=(\lambda_{0k}-\lambda_{1k})^{-1} \sum_{\nu=0}^1 (-1)^{\nu}
\{{\rm exp}(\lambda_{\nu k} \tau_l)
[(D_k+\lambda_{\nu k})f_{k,l-1}+B_kg_{k,l-1}]+$$
\begin{equation}
+\int_{t_{l-1}}^{t_l}{\rm exp}[\lambda_{\nu k}(t_l-\tau)]
[(D_k+\lambda_{\nu k})F'_k(\tau) + B_kG'_k(\tau)]\} d\tau,
\label{app7} 
\end{equation}
$$ g_{k,l}=(\lambda_{0k}-\lambda_{1k})^{-1} \sum_{\nu=0}^1 (-1)^{\nu}
\{{\rm exp}(\lambda_{\nu k} \tau_l)
[(A_k+\lambda_{\nu k})g_{k,l-1}+C_kf_{k,l-1}]+ $$
\begin{equation}
+\int_{t_{l-1}}^{t_l}{\rm exp}[\lambda_{\nu k}(t_l-\tau)]
[(A_k+\lambda_{\nu k})G'_k(\tau) + C_kF'_k(\tau)]\} d\tau
\label{app8}
\end{equation}
where
\begin{equation}
\lambda_{\nu k} =\frac{1}{2} \{-(A_k+D_k)+(-1)^{\nu}[(A_k-D_k)^2
+ 4B_k C_k]^{1/2}\}.
\label{app9}
\end{equation}
Note that for
the  homogeneous environment these relations  are valid for an 
arbitrary value of $\tau_l$, otherwise  they can be applied
only for $\tau_l\ll t_l$.  
  
The knowledge of  $f_{k,0}$ and $g_{k,0}$ (these quantities   
can be calculated on the basis  of boundary conditions 
like  Eq. (\ref{eq41}) or Eq. (\ref{eq42}) ) and the multiple application 
of Eqs (\ref{app7}) -- (\ref{app9}) allow to calculate 
quantities $f_{k,l},g_{k,l}$  for $l=1,2,$ etc.

The   approach of solution of adjoint cascade equations
described above provides an  accuracy better than a 
few per cent and gives results for an arbitrary region 
($\varepsilon_{\rm th}, \varepsilon_{\rm max}$) 
of the primary energy in {\it just} one run of calculations. 
Also  we notice that the computational time consumed 
by this method  does not exceed on average  a few minutes 
for  a  1 GHz  PC type computer and increases  only 
weakly (logarithmically)  with the primary energy. 
In summary, the important feature of this technique 
is its flexibility to describe the cascade processes
in 3 different substances.   It allows large number of calculations
with a good   accuracy throughout  very large 
energy intervals of both primary and secondary energies.
This is an important condition for the quantitative 
analysis and for clear understanding 
of similarities and differences in  cascade development in environments 
dominated by matter, photon gas or   magnetic field.

\section{Elementary processes}

The above described technique requires specification of 
elementary processes that initiate and support   the cascade 
development.  If  we are interested  in the  
longitudinal  development of cascades,
the input should consist of total  cross-sections of interactions  
and  the differential cross-sections as functions of 
energy but   integrated over the emission angles of secondary particles. 
In many astrophysical situations, especially at very high energies,   
this is a fair  approximation, given the  very small (of order 
$\sim m_{\rm e} c^2/E \ll 1$) emission angles of secondary products. 
The one-dimensional  treatment of the cascade development
perfectly works,  if  electrons move along the lines of the 
regular magnetic field. Otherwise, the deviation 
of particles from the cascade core is determined by reflection 
of secondary electrons by magnetic inhomogeneities rather 
than the emission angles. In these cases the diffusion effects must 
be appropriately incorporated  into the cascade equations. 
This question is beyond the present paper.  
 
All processes involved in the electromagnetic cascades
are well known. The cross-sections of these processes  have been  
calculated and comprehensively studied  with very high 
precision  within  the quantum electrodynamics. 

In the case of cascades  in  matter
we take into account the  
ionization losses and bremsstrahlung for electrons 
(positrons), and   the pair production, Compton scattering and 
photoelectric absorption  for photons.  At extremely high energies 
the so-called   Landau-Pomeranchuk-Migdal (LPM)  effect that  
results in suppression of  the bremsstrahlung and pair-production  
cross-sections, may have significant  impact on  the  cascade development. 
The energy  region where the  LPM effect becomes noticeable,  
depends on the atomic number of the ambient matter. 
In the hydrogen dominated  medium this energy region 
appears well beyond $E\ge 10^{20}$~eV, therefore 
in many astrophysical scenarios  the LPM effect 
can be safely ignored.  

Two pairs of coupled processes 
--  the inverse Compton scattering and $\gamma$-$\gamma$ pair production
in the photon gas and the magnetic bremsstrahlung (synchrotron radiation)
and magnetic pair production in the magnetic field --  determine the basic 
features of  cascade produced  in  radiation and magnetic fields.
At extremely high energies the  higher 
order  QED processes may compete with these basic channels. 
Namely,   when the product of 
energies of colliding cascade particles (electrons or photons) 
$E$ and the background photons  $\epsilon$  significantly exceed 
$10^5 - 10^6 m_{\rm e}^2 c^4$, the processes   
$\gamma \gamma \rightarrow e^+ e^-e^+ e^-$\cite{doublpair} 
and  $e \gamma \rightarrow e \gamma e^+e^-$ \cite{triplet}
dominate over the single  ($e^+,e^-$)  pair production and the 
Compton scattering, respectively.  For example, 
in the 2.7 K CMBR the first process stops 
the linear increase  of the mean  free path of highest energy 
$\gamma$-rays  around $10^{21} \ \rm eV$, and puts  
a robust  limit   on the mean free path of $\gamma$-rays 
of about 100 Mpc.  Analogously,  above  $10^{20} \ \rm eV$
the second process  becomes more 
important than the conventional  inverse Compton scattering.
Because  $\gamma \gamma \rightarrow 2e^+ 2e^-$
and   $e \gamma \rightarrow e \gamma e^+ e^-$ channels
result in production of 2 additional  electrons,  they significantly 
change the character of the pure Klein-Nishina pair cascades. However,
an effective  realization of these processes  is possible only at very specific 
conditions with an extremely low  magnetic field and
narrow energy distribution of the background photons. 

Because this study has   methodological objectives,  below  we  do not 
include these  processes  in calculations of cascade characteristics.
This allows us to avoid unnecessary complications and make 
the analysis  more transparent. For the same reason we do not include 
in this study the effect of the  photon splitting \cite{adler} which becomes quite 
important in  pulsars with  magnetic field close to 
$H_{\rm crit}=4.4 \times 10^{13} \ \rm G$ (see e.g. Ref. \cite{HardBaring}).

\subsection{Total cross-sections}

\begin{figure}[htbp]
\centering
\includegraphics[width=0.65\textwidth]{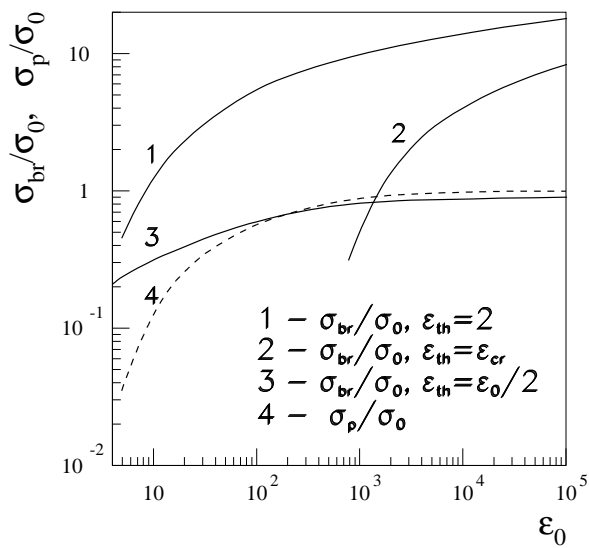} 
\caption{Total cross-sections 
of the bremsstrahlung ($\sigma_{\rm br}$) and pair production 
($\sigma_{\rm p}$) processes in hydrogen gas normalized to the 
asymptotic value ($\sigma_0$) of the pair production cross-section 
at  $\varepsilon_0\to \infty$.
The bremsstrahlung cross-sections are calculated for secondary 
\grs produced with energies exceeding 
(1) the pair-production threshold,  $\varepsilon_{\rm th}=2$;
(2) the critical energy,  $\varepsilon_{\rm th} \simeq 700$;
(3) half of the energy of the primary electron,
$\varepsilon_{\rm th} = \varepsilon_0/2$.
}
\label{fgr1}
\end{figure}

\begin{figure}[htbp]
\centering
\includegraphics[width=0.65\textwidth]{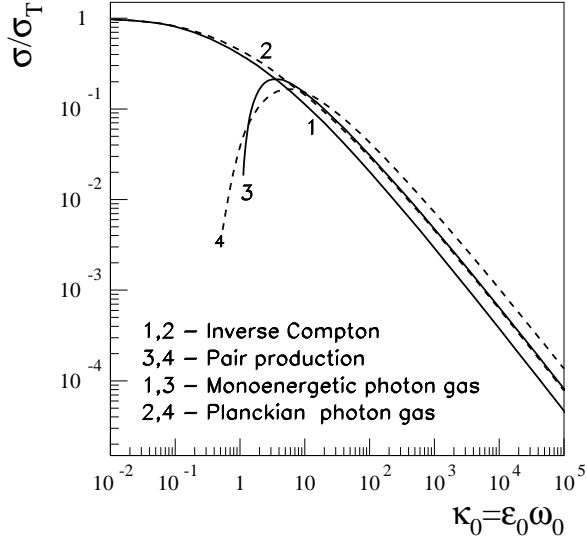} 
\caption{Total cross-sections of the inverse Compton scattering 
and pair production.  Two spectral distributions 
for the ambient photon  gas are assumed - 
mono-energetic  with energy 
$\omega_0$ (curves 1 and 3) and Planckian  with 
the same mean photon energy $\omega_0 \simeq 3 k T/m_{\rm e}c^2$.}
\label{fgr2}
\end{figure}

\begin{figure}[htbp]
\centering
\includegraphics[width=0.65\textwidth]{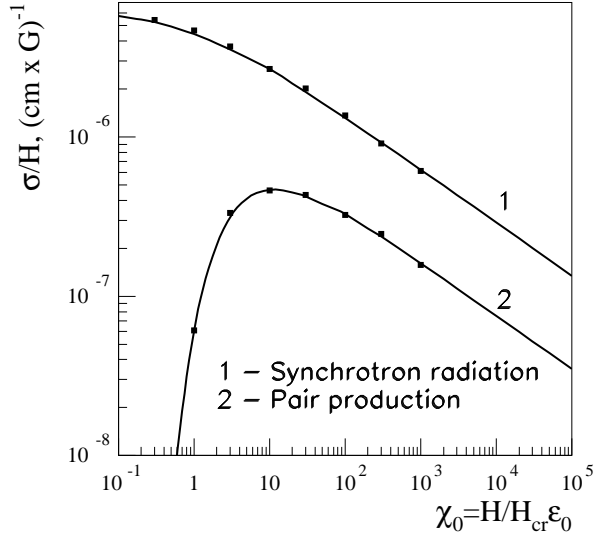} 
\caption{Total cross-sections of the synchrotron radiation and 
magnetic pair production. 
Solid curves -- our calculations; points -- calculations of 
Anguelov et.al. \cite{Anguelov}.}
\label{fgr3}
\end{figure}

In Figs.~\ref{fgr1},\ref{fgr2} and \ref{fgr3} we show 
the photon-  and pair-production  
cross-sections in hydrogen gas and in  radiation  and magnetic 
fields,  respectively. The   energies of  electrons and  $\gamma$-rays 
are expressed in units  of $m_{\rm e} c^2$. 

% Figure 1.
% ---------------

In Fig.~\ref{fgr1} the total cross-sections for matter are presented.
They are normalized to  the asymptotic value of the pair production 
cross-section at $\varepsilon_0\to \infty$:   
\begin{equation}
\sigma_0= 7/9\times 4\alpha_{\rm f}   r_{\rm e}^2 Z(Z+1) 
\frac{\ln (183 Z^{-1/3})}{1+0.12 (Z/82)^2}
\label{radlmatter}
\end{equation}
where $Z$ is the charge of the target nucleus, $r_{\rm e}$ is the 
classical  electron radius. This actually implies introduction of the 
so-called radiation length
\begin{equation}
X_0^{(M)}= 7/9 \left[ n_0^{(M)} \sigma_0\right]^{-1}
\end{equation}
 
$X_0^{(M)}$ has a meaning of the average distance 
over  which the  ultrarelativistic electron loses  all but 1/e of 
its energy due to bremsstrahlung. The same parameter
approximately corresponds to the mean free path of  $\gamma$-rays.
Therefore, the cascade effectively  develops 
at depths exceeding the radiation length. Usually the radiation 
length  is expressed in units $\rm g/cm^2$. For the hydrogen gas 
$X_0^{(M)}=63 \ \rm g/cm^2$.  The second important parameter that    
characterizes  the cascade development 
is the so-called {\rm critical energy}  below which the ionization 
energy losses dominate over  bremsstrahlung losses. 
In the hydrogen gas    
$ \varepsilon_{\rm cr}\simeq 700 $.
Effective  multiplication of particles due to the cascade processes
is possible only at energies 
$\varepsilon\ge \varepsilon_{\rm cr}$.
At lower energies electrons dissipate their energy 
by ionization rather than producing more high energy \grs 
which would support further development of the electron-photon shower.  

The bremsstrahlung differential  cross-section   
has a $1/\varepsilon_\gamma$ type 
singularity  at  $\varepsilon_{\gamma}\to 0$ 
($\varepsilon_{\gamma}$ is the energy of the emitted photon),
but because of  the hard spectrum of bremsstrahlung photons 
the energy losses of electrons are contributed 
mainly by emission of high energy $\gamma$-rays.  
In Fig.~\ref{fgr1} we show the  bremsstrahlung total cross-sections
calculated for  3  different values of minimum energy of 
emitted photons:  
$\varepsilon_{\rm th} =2$, $\varepsilon_{\rm cr}$ and 
$\varepsilon_{\rm e}/2$. The first value corresponds to the 
cross-section of production of all \grs capable to produce 
electron-positron pairs. The second value corresponds to the 
cross-section of production of \grs produced above the critical energy,
and thus capable to  support the cascade.
And finally, the third value corresponds  to the 
cross-section of production of the ``most important'' 
\grs which play the major role 
in the cascade development. 
It is seen from Fig.~\ref{fgr1}  that  while 
for $\varepsilon_{\rm th} =2$
the total cross-section of pair-production is  an order
of magnitude lower compared to the bremsstrahlung cross-section,
for $\varepsilon_{\rm th} = \varepsilon_{\rm e}/2$ the 
cross-sections of two processes become almost identical 
at energies $\varepsilon \geq  100 $.

% Figure 2.
% ----------------------------------------------------------------------------

Both cross-sections  grow  significantly with  energy until 
$\varepsilon_0 \sim 10^3$. At higher   energies the 
pair-production cross-section   becomes energy-independent,
but for $\varepsilon_{\rm th} =2$ the bremsstrahlung cross-section 
continues to grow, although very  slowly  (logarithmically).

% Table 1. 
% -----------------------------------------------------------
\begin{table}
\caption{The asymptotic behaviour of the total cross-sections 
of electron and photon interactions  at high energies.}
\bigskip
\begin{center}
\begin{tabular}{lccc} 
\hline
  Environment  & Matter  & Photon gas & Magnetic field \\ \hline
  $\sigma_e$   &$\sim {\rm ln}\varepsilon_0$ & 
$\sim (\kappa_0\cdot {\rm ln}\kappa_0)^{-1}$&
$\sim\chi_0^{-1/3}$ \\ 
  $\sigma_\gamma$   &$\sim {\rm const}$ & 
$\sim (\kappa_0\cdot {\rm ln}\kappa_0)^{-1}$&
$\sim\chi_0^{-1/3}$ \\ \hline
 \end{tabular}
\end{center}
\end{table}

In Fig. \ref{fgr2} we show the total cross-sections of 
the inverse Compton scattering and pair-production 
for the mono-energetic and Planckian isotropic photon fields, normalized 
to the  Thompson cross section 
$\sigma_{\rm T}=8/3 \pi r_{\rm e}^2=6.65 \times 10^{-25} \ \rm cm^2$.
Both cross sections depend on the product 
$\kappa_0 = \varepsilon_0 \omega_0$ of the primary ($\varepsilon_0$)
and ambient ($\omega_0=\epsilon/m_{\rm e} c^2$) photon energies.
  At $\kappa_0 \to 0$ the inverse
Compton cross-section   $\sigma_{\rm IC} \to \sigma_{\rm T}$. 
At high energies  it decreases   with $\kappa_0$, as 
$\propto  (\kappa_0\cdot{\rm ln}\kappa_0)^{-1}$.

% Figure 3.
% ----------------------------------------------------------------------------

In   the mono-energetic radiation field the pair production has a 
strict  kinematic  threshold at    $\kappa_0=1$. At small values 
of  $\kappa_0 \leq 2$ the cross-section rapidly increases with
 $\kappa_0$ achieving its maximum of about $\simeq 0.2 \sigma_{\rm T}$
at $\kappa_0 \simeq 3.5-4$, and decreases with further increase of 
$\kappa_0$ as $\propto (\kappa_0\cdot{\rm ln}\kappa_0)^{-1}$. 
Thus in this regime the pair-production cross-section  
behaves quite similar to  the Compton scattering in the Klein-Nishina 
regime, but its absolute value  is higher by a factor of 1.5.     
    
In the case of interactions of electrons and photons 
with the magnetic field it is more  
convenient  to introduce,  instead of standard total 
cross-sections, the  interaction  probabilities \cite{Anguelov}.
But in the literature this parameter is still formally called as 
cross-section.  
These probabilities normalized to the strength of the magnetic field 
are shown in Fig. ~\ref{fgr3}. The probabilities of both processes --  
photon production by electrons (synchrotron radiation) 
and pair production by photons  -- depend only on the  
parameter $\chi_0= \varepsilon_0 H/H_{\rm cr} $, 
where H is the component 
of magnetic field intensity perpendicular to the velocity
of the particle.  
The parameter  $\chi_0=H/H_{\rm cr}\varepsilon_0$ 
is an apparent  analog of  the parameter $\kappa_0$. 
While the probability of the synchrotron radiation
at   $\chi_0 \ll 1$ is constant, the probability of the pair 
production drops  dramatically below  $\chi_0= 1$. 
After achieving its maximum at $\chi_0\simeq 10$, the probability of 
the  pair production decreases with $\chi$ as 
$\propto \chi_0^{-1/3}$. The same behaviour at large
 $\chi$ has also the probability of synchrotron radiation, but 
the absolute value of the latter always exceeds by a factor of 3 the 
probability of the pair production.

% Table 2. 
% -----------------------------------------------------------
\begin{table}
\caption{The singularities in the  differential 
cross-sections of cascade processes.}
\bigskip 
\centering
\begin{tabular}{lcc} 
\hline
 Elementary process     & Singularity point& Behaviour \\ \hline
 Bremsstrahlung          & $\varepsilon_{\gamma} = 0$& 
  $\sim \varepsilon_{\gamma}^{-1}$ \\ 
 Inverse Compton scattering& $\varepsilon_{\gamma}= \varepsilon_0$ & 
  $\sim (\varepsilon_0-\varepsilon_{\gamma})^{-2}$ \\  
 Pair production in photon gas& $\varepsilon_e=0$ &
$\sim \varepsilon_e^{-2}$ \\ 
 Pair production in photon gas & $\varepsilon_e=\varepsilon_0$ & 
  $\sim (\varepsilon_0-\varepsilon_e)^{-2}$ \\ 
  Synchrotron radiation & $\varepsilon_{\gamma}=0$  &
  $\sim \varepsilon_{\gamma}^{-2/3}$ \\ \hline
\end{tabular}
\end{table}

% Table 3. 
% -----------------------------------------------------------
\begin{table}
\caption{The mean portion of primary energy 
($\bar{\varepsilon}_{\gamma}/\varepsilon_0$) transferred to the secondary 
photon in the inverse Compton scattering (ics) and the synchrotron 
radiation (syn) processes via the  values of parameters 
$\kappa_0=\varepsilon_0\omega_0$ and 
$\chi_0=\varepsilon_0 H/H_{\rm cr}$, respectively.}
\bigskip
\centering
\begin{tabular}{lcccccc} 
\hline
$\kappa_0$, $\chi_0$&                             
0.01 & 0.1 &    1& $10^2$& $10^4$ &$10^6$ \\ \hline
$(\bar{\varepsilon}_{\gamma}/\varepsilon_0)_{\rm ics}$&
0.014&0.099&0.358&  0.760& 0.867& 0.910   \\    
($\bar{\varepsilon}_{\gamma}/\varepsilon_0)_{\rm syn}$&
$0.44\cdot 10^{-2}$& 
0.033&0.118& 0.241&  0.250& 0.250   \\ \hline     
\end{tabular}
\end{table}

% Table 4. 
% -----------------------------------------------------------
\begin{table}
\caption{The mean portion of primary energy 
($\bar{\varepsilon}_e/\varepsilon_0$) transferred to the 
leading secondary particle for the processes of pair production 
in a mono-energetic photon gas ($G$) and in the magnetic field ($F$) 
via the  values of parameters 
$\kappa_0=\varepsilon_0\omega_0$ and 
$\chi_0=\varepsilon_0 H/H_{\rm cr}$, respectively.}
\bigskip
\centering
\begin{tabular}{lcccccc} 
\hline
$\kappa_0$, $\chi_0$&                             
1 & 3 &    10& $10^2$& $10^4$ &$10^6$ \\ \hline
$(\bar{\varepsilon}_e/\varepsilon_0)_{\rm G}$&
0.500&0.701&0.797&  0.891& 0.948& 0.966   \\ 
($\bar{\varepsilon}_e/\varepsilon_0)_{\rm F}$&
0.634&0.693& 0.746&  0.782& 0.824& 0.825   \\ \hline     
\end{tabular}
\end{table}

% Figure 4.
% ----------------------------------------------------------------------------
\begin{figure}[htbp]
\begin{center}
\includegraphics[height=0.30\textheight]{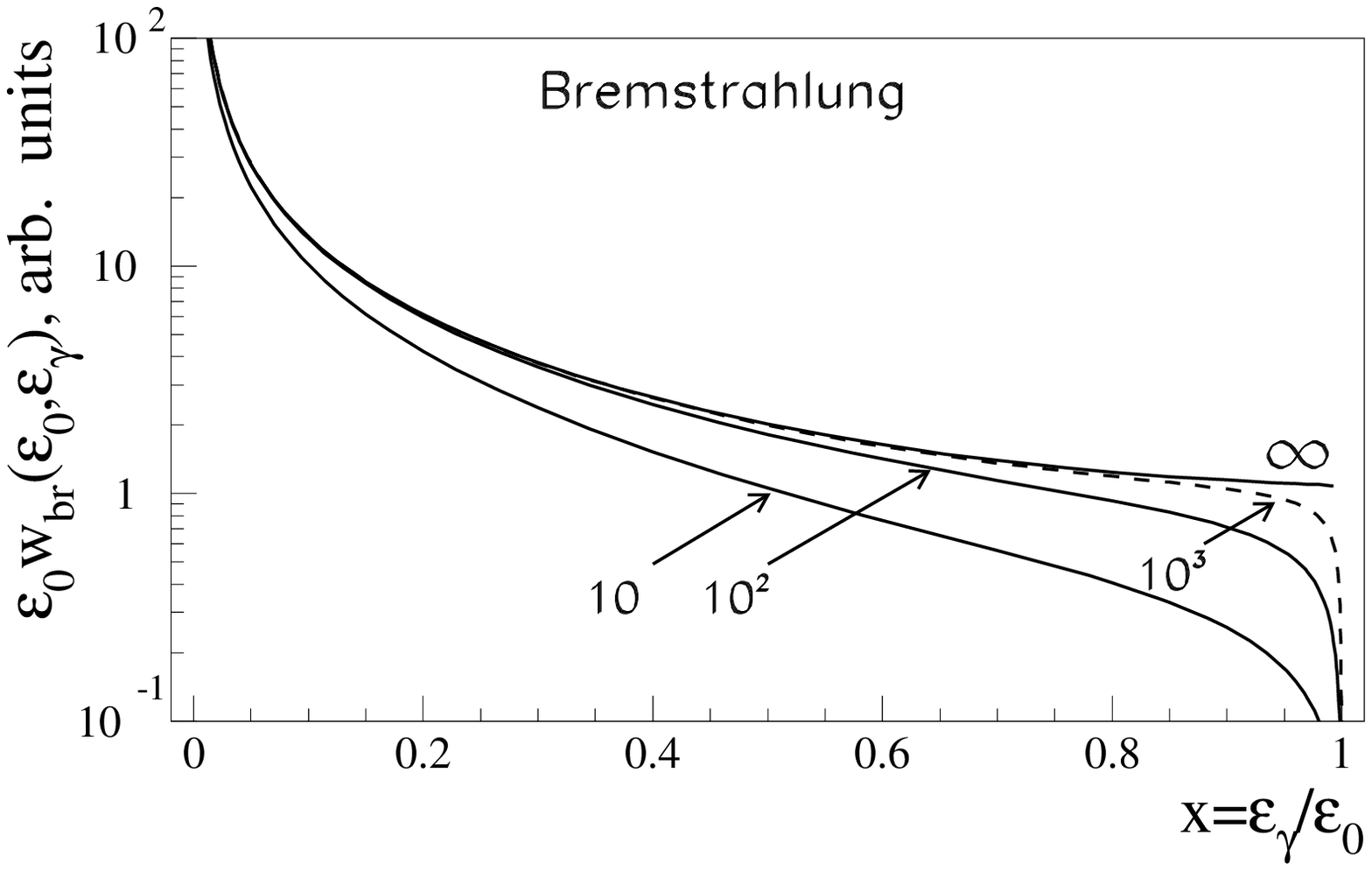}
\includegraphics[height=0.30\textheight]{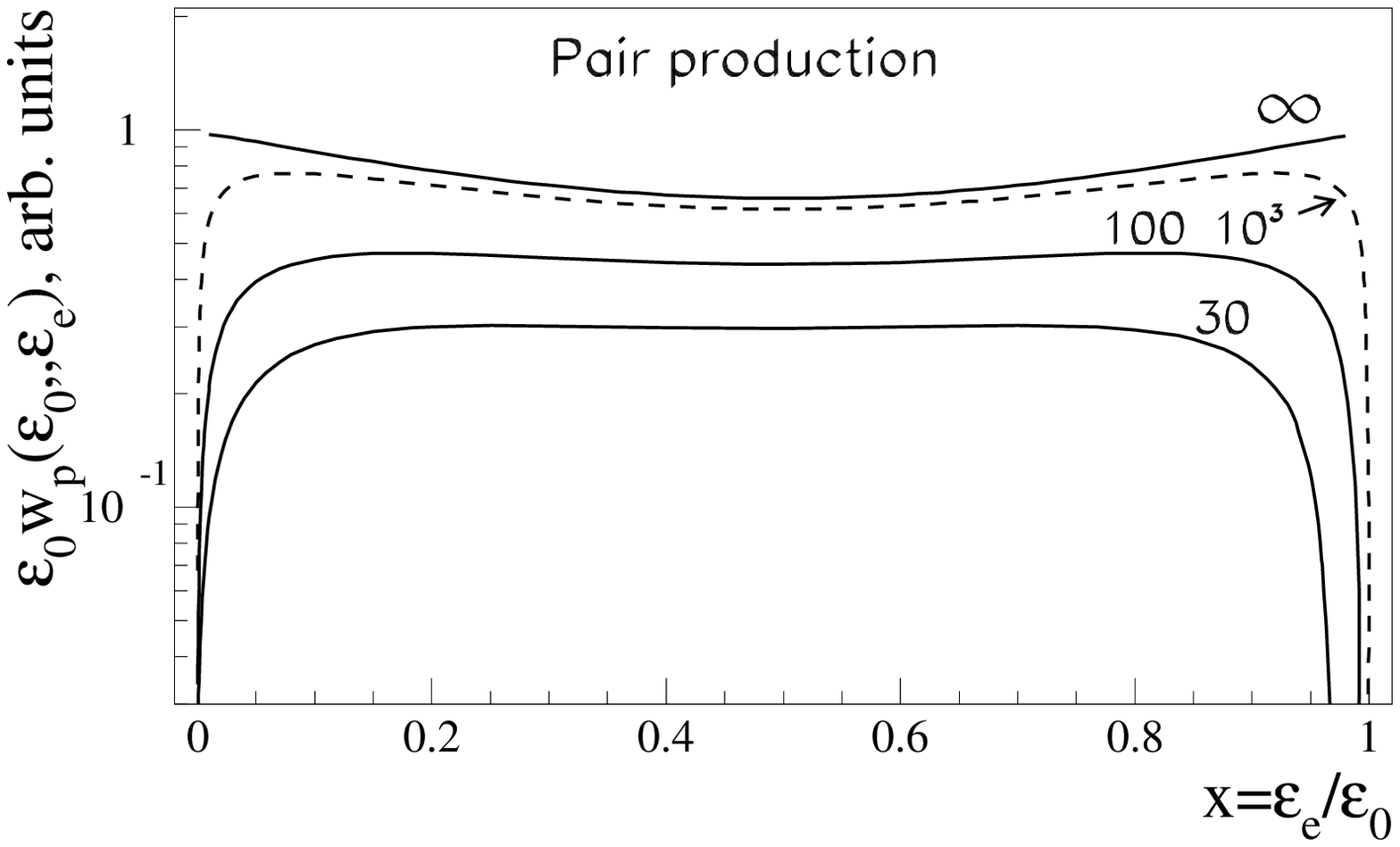}
\end{center} 
\caption{Differential  cross-sections of the bremsstrahlung 
(upper panel) and  pair production (bottom panel) processes  in hydrogen.
The cross-sections  are normalized to 
one radiation length.  The energies of primary electrons and \grs  
$\varepsilon_0$ are indicated at the curves.}
\label{fgr4}
\end{figure}

% Figure 5.
% ----------------------------------------------------------------------------
\begin{figure}[htbp]
\begin{center}
\includegraphics[height=0.30\textheight]{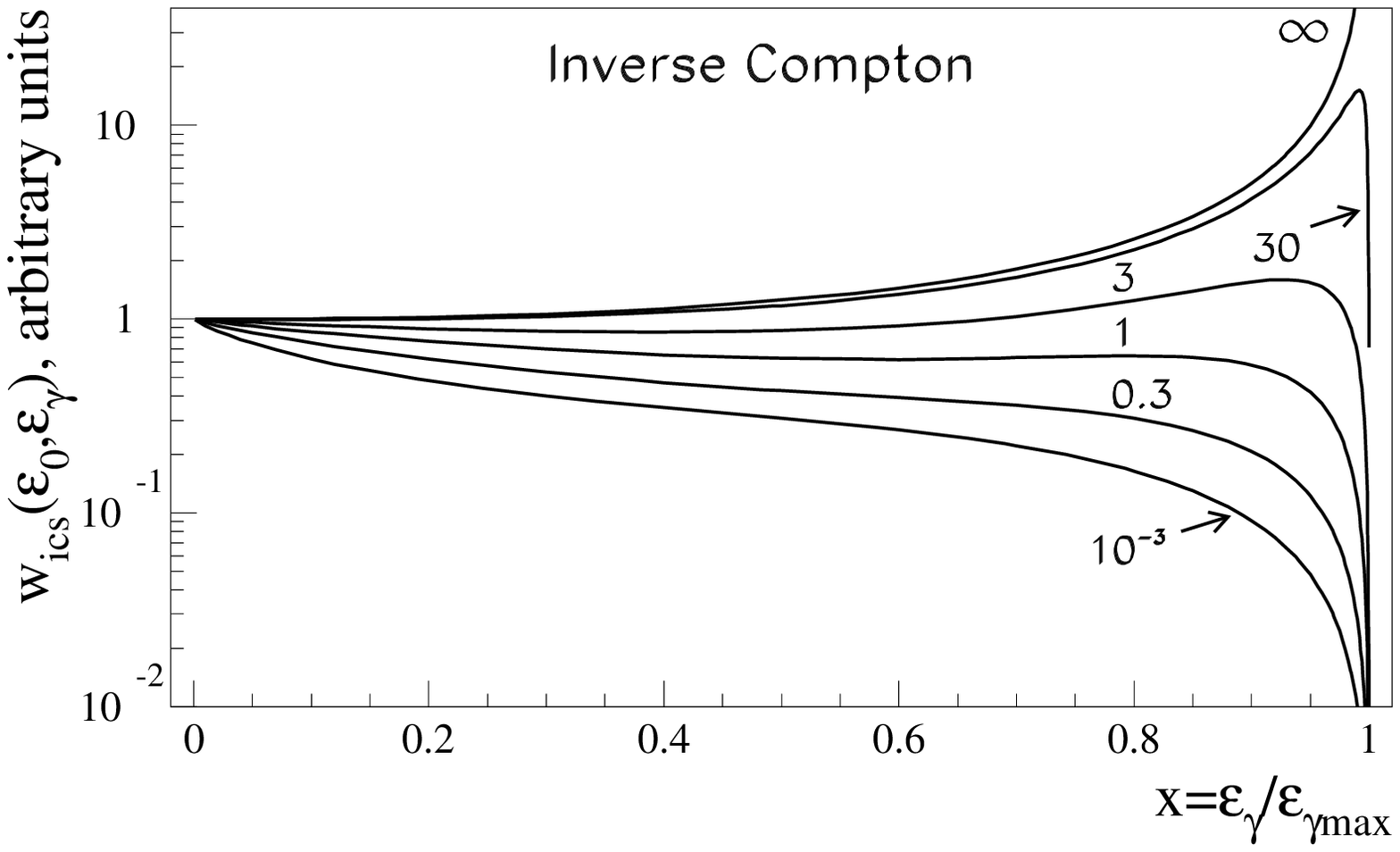}
\includegraphics[height=0.30\textheight]{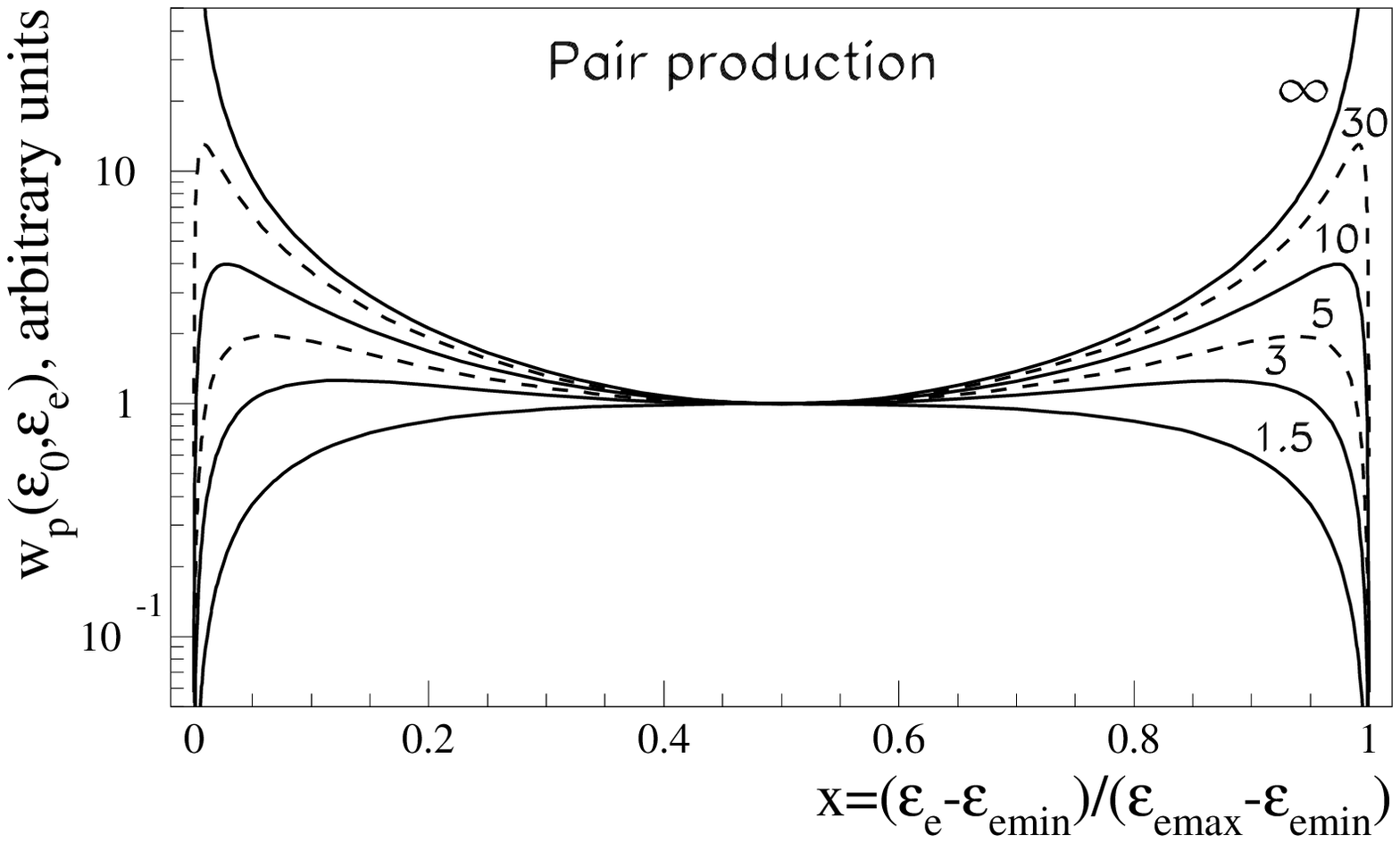}
\end{center}
\caption{Differential  cross-sections of the inverse Compton 
(upper panel) and pair production (bottom panel) processes 
for the case  of a mono-energetic gas of ambient photons.
The parameters $\varepsilon_{\gamma max}$, $\varepsilon_{\rm e min}$
and $\varepsilon_{\rm e max}$  are defined as 
$\varepsilon_{\gamma \rm max}= 4 \varepsilon_0 (\kappa_0/1+4\kappa_0 )$ and 
$\varepsilon_{\rm e min,\, e max}=0.5 \varepsilon_0 
(1 \mp \sqrt{1-1/\kappa_0})$.
Different values of the parameter  $\kappa_0=\varepsilon_0\omega_0$ 
are indicated at the  curves.} 
\label{fgr5}  
\end{figure}

% Figure 6.
% ----------------------------------------------------------------------------
\begin{figure}[htbp]
\begin{center}
\includegraphics[height=0.30\textheight]{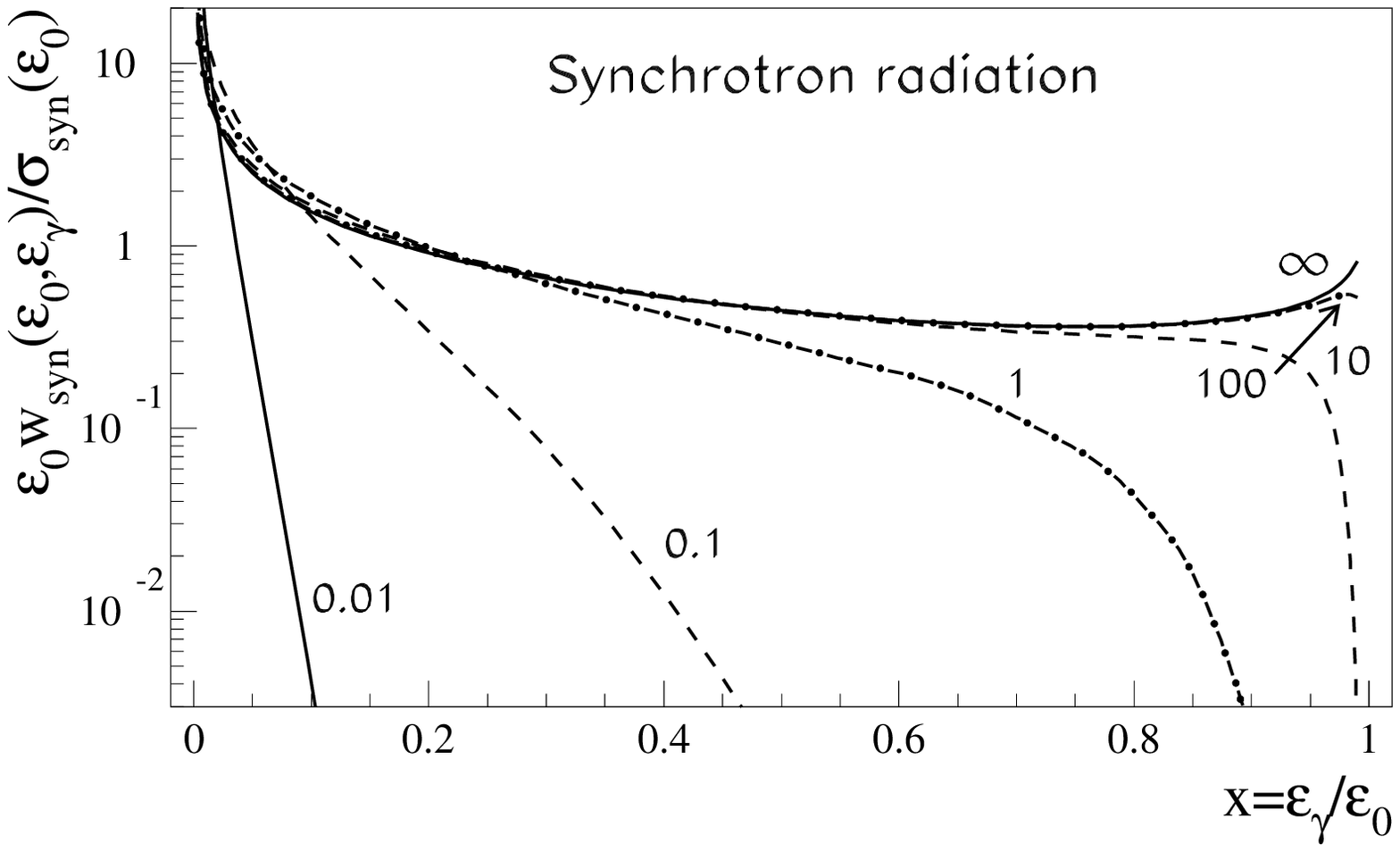}
\includegraphics[height=0.30\textheight]{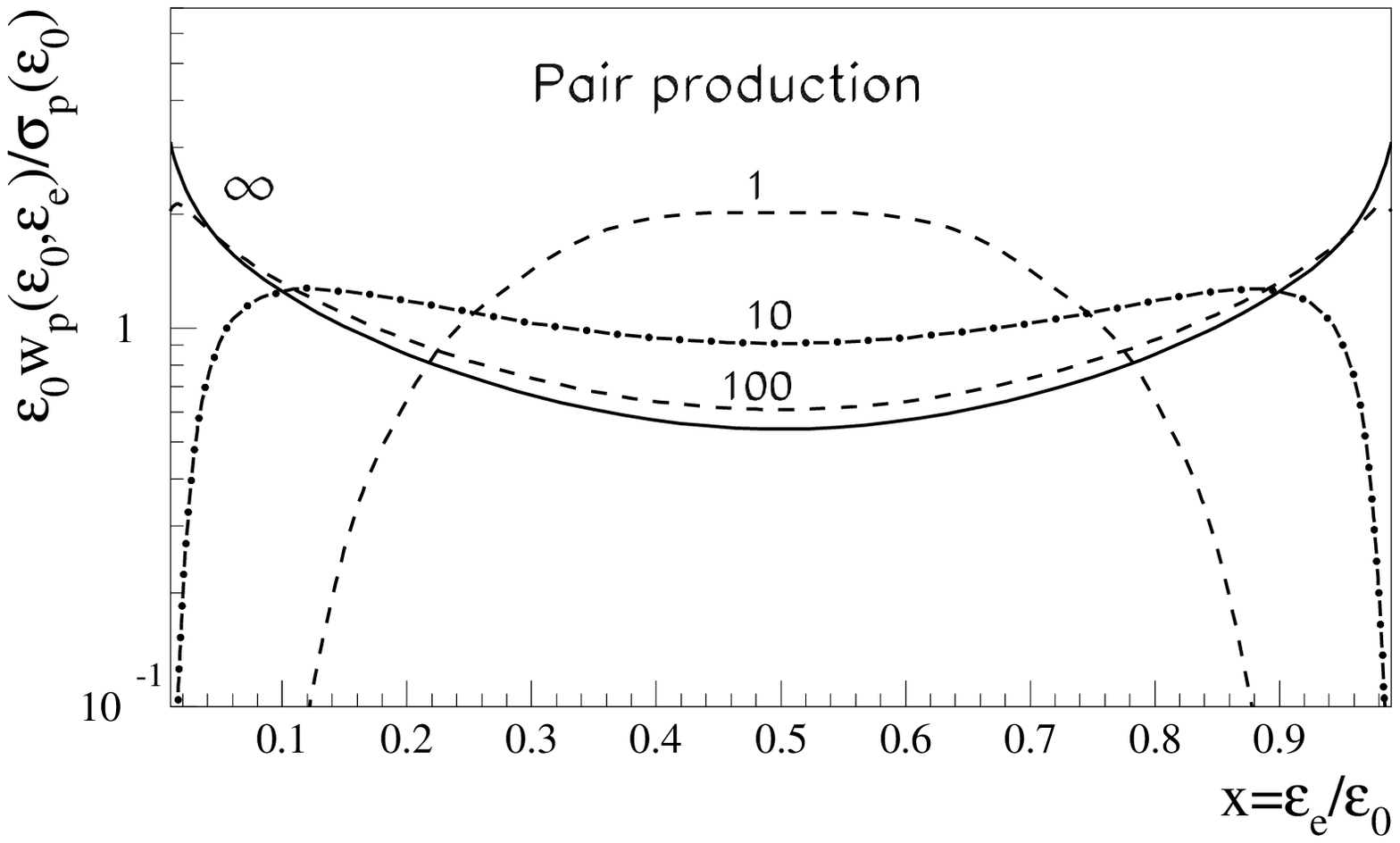}
\end{center}
\caption{Differential cross-sections  of synchrotron radiation 
(upper  panel) and the magnetic pair production (bottom panel) 
normalized to the total cross-sections 
of these processes.  Different values of parameter 
$\chi_0=H/H_{\rm cr}\varepsilon_0$ are  indicated at the curves.}
\label{fgr6}
\end{figure}

\subsection{Differential cross-sections of cascade processes}

The differential cross-sections of cascade processes 
are presented  in Figs.~\ref{fgr4},\ref{fgr5} and \ref{fgr6}. 
The bremsstrahlung 
and pair-production cross-sections are  
from  Ref. \cite{Akh_Berest,BlumGould}. The cross-sections for the 
inverse Compton  scattering and pair production in the mono-energetic 
isotropic photon field  are  from Ref. \cite{BlumGould} and 
Ref. \cite{AhAtNah}, respectively. The cross-sections of processes 
in the magnetic field  are from Ref. \cite{Akhiezer}. 

All three cross-sections of pair-production 
are symmetric functions in  respect to the point
$x=\varepsilon_e/\varepsilon_0=0.5$. The photon 
production processes are asymmetric functions.  
The bremsstrahlung, synchrotron radiation, the  
inverse Compton scattering and the pair production  
in the photon gas  have  singularities. The location of singularities
and  the cross-section  behaviour near the singularity points 
are summarized in Table~2.

The character of cascade development is largely determined by the 
fraction of energy of electrons and photons  lost  per interaction.  
In Table~3 we present    mean fractions  of the electron energy  
transferred to the secondary photons in the 
inverse Compton and synchrotron radiation processes
at different values of $\kappa_0$ and  $\chi_0$. 
In the classical regime  ($\kappa_0\ll 1$ and  $\chi_0\ll 1$)
the mean energy lost by the electron 
per interaction is very small, $\Delta \varepsilon /\varepsilon \ll 1$.
In the "quantum"  regime  ($\kappa_0\gg 1$ 
and  $\chi_0\gg 1$)   the interactions have a catastrophic 
character; the secondary  photons get a significant fraction of 
the energy of parent electrons.  In the photon gas  
at $\kappa_0 \gg 1$  this fraction exceed 0.5, 
approaching asymptotically to 1. In the magnetic field the  
energy transfer is smaller; at $\chi_0\sim 1$
it is approximately 0.1, and asymptotically approaches  
to 1/4  at extremely large $\chi_0$. 

The pair production processes in all 3 substances
have (by definition)  catastrophic character (the photon 
always disappears). Since the differential spectra of secondary electrons 
are quite flat with increase towards the maximum energy 
$\varepsilon_{\rm e}  \to  \varepsilon_\gamma$ (in the 
ultrarelativistic regime),  these processes proceed 
with formation of leading electron with energy 
$x=\varepsilon_e/\varepsilon_\gamma \to 1$ 
in  the photon gas and somewhat smaller 
($\simeq 0.8$)  in  the magnetic field (see Table 4).

\section{Cascades}

In this section we discuss and compare the 
so-called cascade curves and the energy spectra of 
electrons and photons  for showers  produced 
in  matter,  photon gas and magnetic field.  

\subsection{Cascade curves}

The cascade curve  
$N_{\beta}(t,\varepsilon_0,\varepsilon_{\rm th})$ describes the 
average number of cascade electrons   ($\beta=e$)  or photons 
($\beta=\gamma$) above $\varepsilon_{\rm th}$,  
as a function of   the penetration depth $t$.

\subsubsection{Matter}

\begin{figure}[htbp]
\centering
\includegraphics[width=0.65\textwidth]{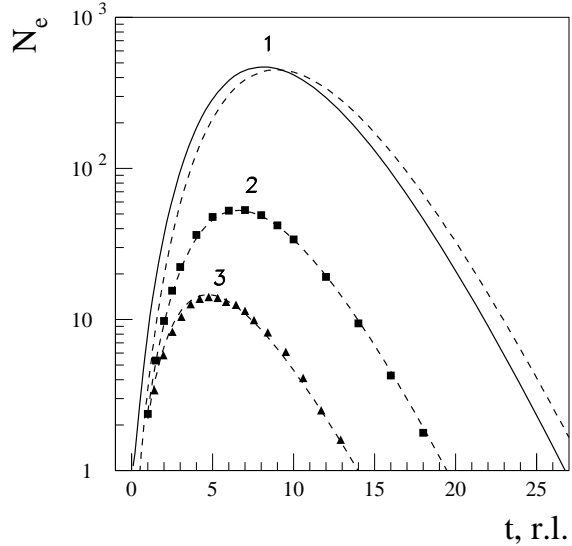} 
\caption{Cascade curves of electrons for showers initiated 
by primary electrons (solid curves) 
and photons (dashed curves). The calculations are performed for the 
following primary energies 
$\varepsilon_0=2\cdot 10^8$~(curve 1),
$2\cdot 10^7$~(curve 2), $2\cdot 10^4$~(curve 3) 
and the ratio   $\varepsilon_{th}/\varepsilon_{cr}= 125$~(curves 1 and 2),
$0.05$~(curve 3).   For comparison,  
the results  derived from  the analytical  
cascade  theory \cite{matter} (boxes) and 
by simulations with the ALTAI code \cite{Altai} (triangles)
are also shown}
\label{fgr7}
\end{figure}

In Fig.~\ref{fgr7} we show  the cascade curves  for electrons calculated  
for  different values  of the primary ($\varepsilon_0$) and 
threshold ($\varepsilon_{\rm th}$) energies using 
the adjoint equation technique.  For comparison we also show  
the cascade curves calculated  within the  approximation A of 
the analytical theory of electromagnetic cascades \cite{matter}  
(valid for  $\varepsilon_{\rm th}\gg \varepsilon_{\rm cr}$),  
as well as  the cascade curves  obtained by  Monte Carlo simulations 
using the ALTAI code\cite{Altai}.  The results obtained 
by 3 different methods are in a good agreement with each other. 

One of the basic parameters characterizing the cascade development 
is the depth $t_{\rm m}$   (expressed in units of radiation 
length $X_0^{(M)}$) 
at which  the cascade curve achieves its  maximum.  
Generally  $t_{\rm m}$  grows  logarithmically with the 
primary energy. In particular, for light materials  (with 
the nucleus charge number 
$Z \le 10$) and for $\varepsilon_0 \gg \varepsilon_{\rm cr}$, the
parameter   $t_{\rm m}$ for electrons of a cascade initiated by 
a primary photon can be approximated as \cite{matter}:
\begin{equation}
t_{\rm m}^{\rm e} \simeq \mbox{ln} (\varepsilon_0/\varepsilon_{\rm th})
\,\, \mbox{for} \,\, \varepsilon_{\rm th}\gg \varepsilon_{\rm cr}; \quad
t_{\rm m} \simeq \mbox{ln}(\varepsilon_0/\varepsilon_{\rm cr}) \,\,
\mbox{for} \,\, \varepsilon_{\rm th}\to 0. 
\end{equation}
The primary electron interacts with
matter somewhat earlier than the primary $\gamma$-ray
(see Fig.~\ref{fgr1}), therefore
$t_{\rm m}^\gamma \approx t_{\rm m}^{\rm e}+0.7$. 

Another important parameter is  the total number of electrons 
at the shower maximum:
$N_{\rm max}=N_{\rm e}(t_{\rm m},\varepsilon_0,\varepsilon_{\rm th})$.
This number is approximately proportional to the primary energy.
In the energy region   $\varepsilon_{\rm th}\gg \varepsilon_{\rm cr}$
the electron number $N_{\rm em}\sim \varepsilon_{\rm th}^{-1}$; however it 
does not depend on $\varepsilon_{\rm th}$, if $\varepsilon_{\rm th}\to 0$. 

\begin{figure}[htbp]
\centering
\includegraphics[width=0.65\textwidth]{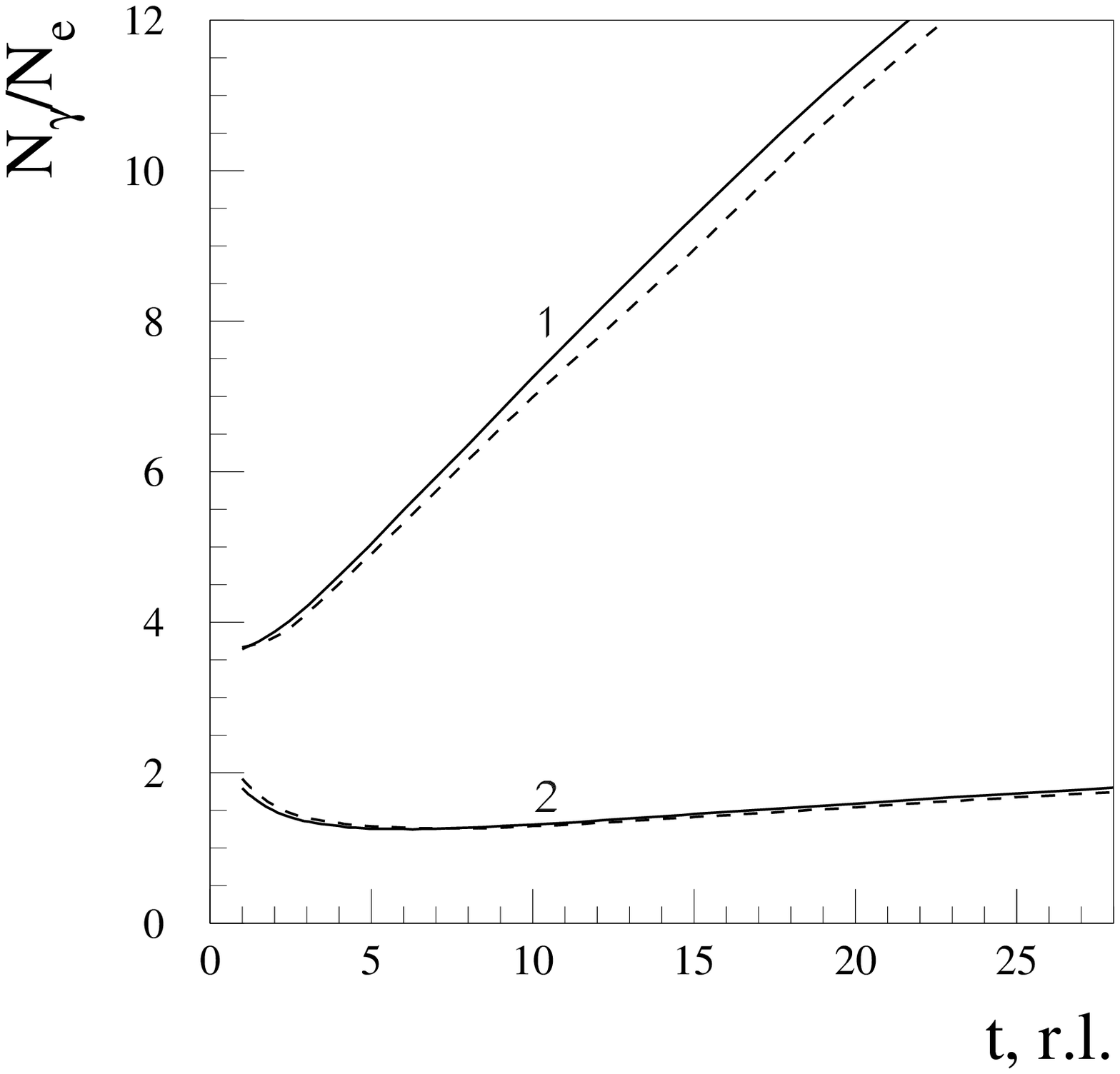}
\caption{The depth dependence of the $N_{\gamma}/N_e$ ratio for 
cascades initiated by primary electrons (solid curves) 
and photons (dashed curves). Primary energy
$\varepsilon_0=2\cdot 10^8$; 
$\varepsilon_{th}/\varepsilon_{cr}=0$~(curves 1), $125$~(curves 2).}
\label{fgr8}
\end{figure}

In Fig.~\ref{fgr8} we present the ratio of the number of the cascade
photons  to  electrons as  a function of the penetration depth. 
For  $\varepsilon_{\rm th}\gg \varepsilon_{\rm cr}$,  the 
number of cascade photons $N_{\gamma}$ is comparable to  the electron 
number $N_{\rm e}$.  At  small threshold energies,    
($\varepsilon_{\rm th} \ll  \varepsilon_{\rm cr}$) the number of $\gamma$-rays 
considerably exceeds the number of electrons. This is  explained by the
break of symmetry between the electrons and  $\gamma$-rays caused by 
ionization losses of electrons.

% Figure 7.
% ----------------------------------------

% Figure 8.
% ----------------------------------------------------------------------------

\subsubsection{Photon gas}

For description of the cascade development in the mono-energetic
photon gas it is convenient to introduce, 
analogously  to the cascade in matter, the 
radiation length $X_0^{\rm (G)}$ in the following form
\begin{equation}
X_0^{\rm (G)}=\left[ 4\pi  n_0^{\rm (G)} r_{\rm e}^2
\right]^{-1} \kappa_0 \ ,
\label{rlphoton}
\end{equation} 
where $n_0^{\rm (G)}$ is the number density of photons.
Apparently, $X_0^{\rm (G)}$ corresponds,
with accuracy up to logarithmic terms, 
 to the mean free path 
of $\gamma$-rays  in the photon field  at $\kappa_0 \gg 1$.  

Note that unlike to the radiation length in matter, 
the primary energy  enters,  through the parameter 
$\kappa_0 = \varepsilon_0 \omega_0$, in  $X_0^{\rm (G)}$. 
Obviously, for the ambient photon gas with a broad 
band energy distribution this parameter becomes meaningless. 
At the same time in a narrow band radiation fields, e.g. with Planckian
distribution, this parameter can work effectively by substituting 
$\bar{\omega}_0 \approx 3 kT/ m_{\rm e} c^2$. Note that 
for the black-body radiation the  density of photons is determined by 
temperature, therefore
\begin{equation}
X_0^{\rm (BB)}= 3/\pi \alpha_{\rm f}^{3} r_0  
(kT/m_{\rm e}c^2)^{-2} \varepsilon \simeq
6.9 \times 10^{-7} (kT/m_{\rm e}c^2)^{-2} \varepsilon \  \rm cm
\label{rlbb}
\end{equation} 
In particular,  in the 2.7 K CMBR 
$X_0^{\rm (BB)} \simeq 3.3 \times 10^{12} \varepsilon \ \rm cm$.  

\begin{figure}[htbp]
\centering
\includegraphics[height=0.30\textheight]{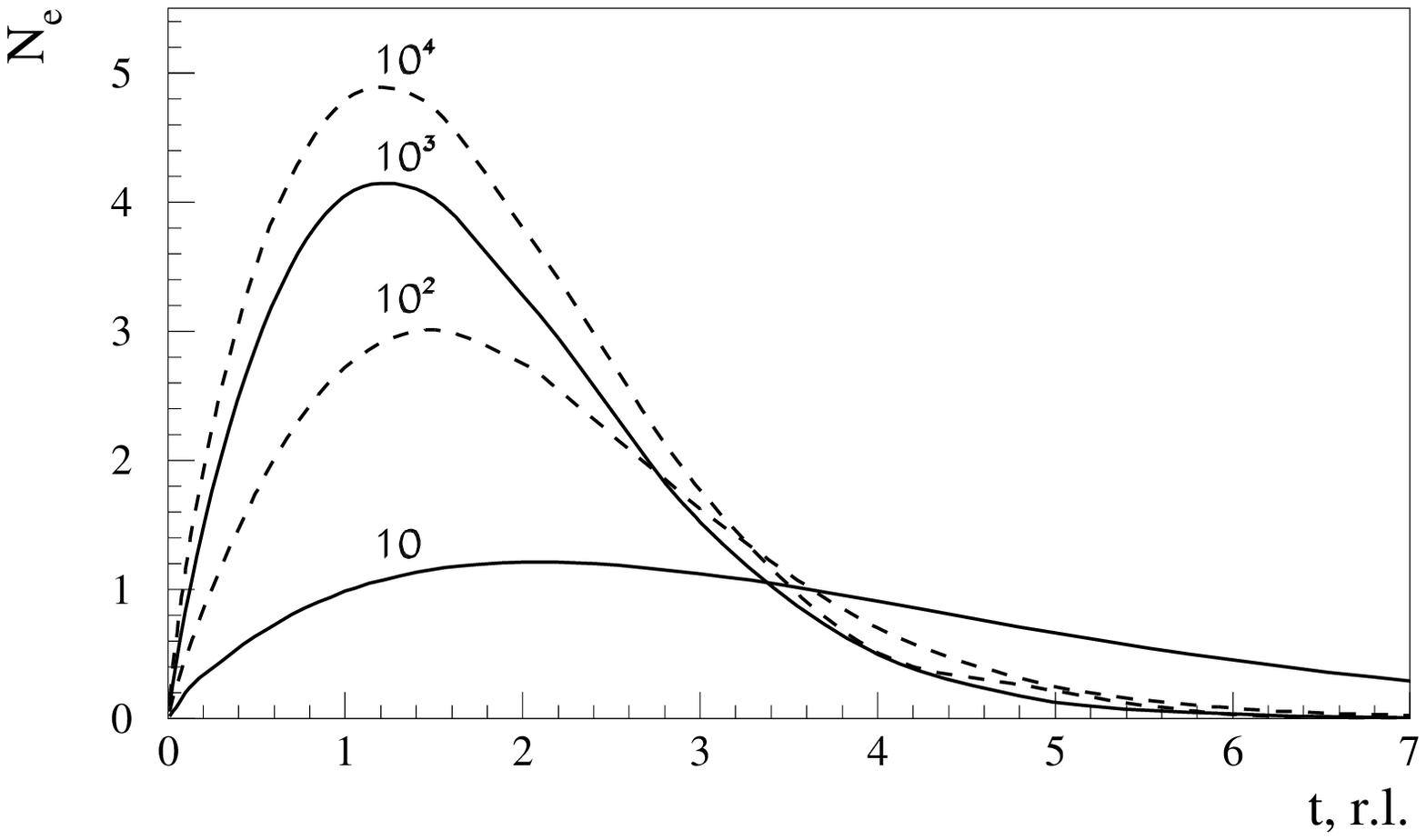} 
\includegraphics[height=0.30\textheight]{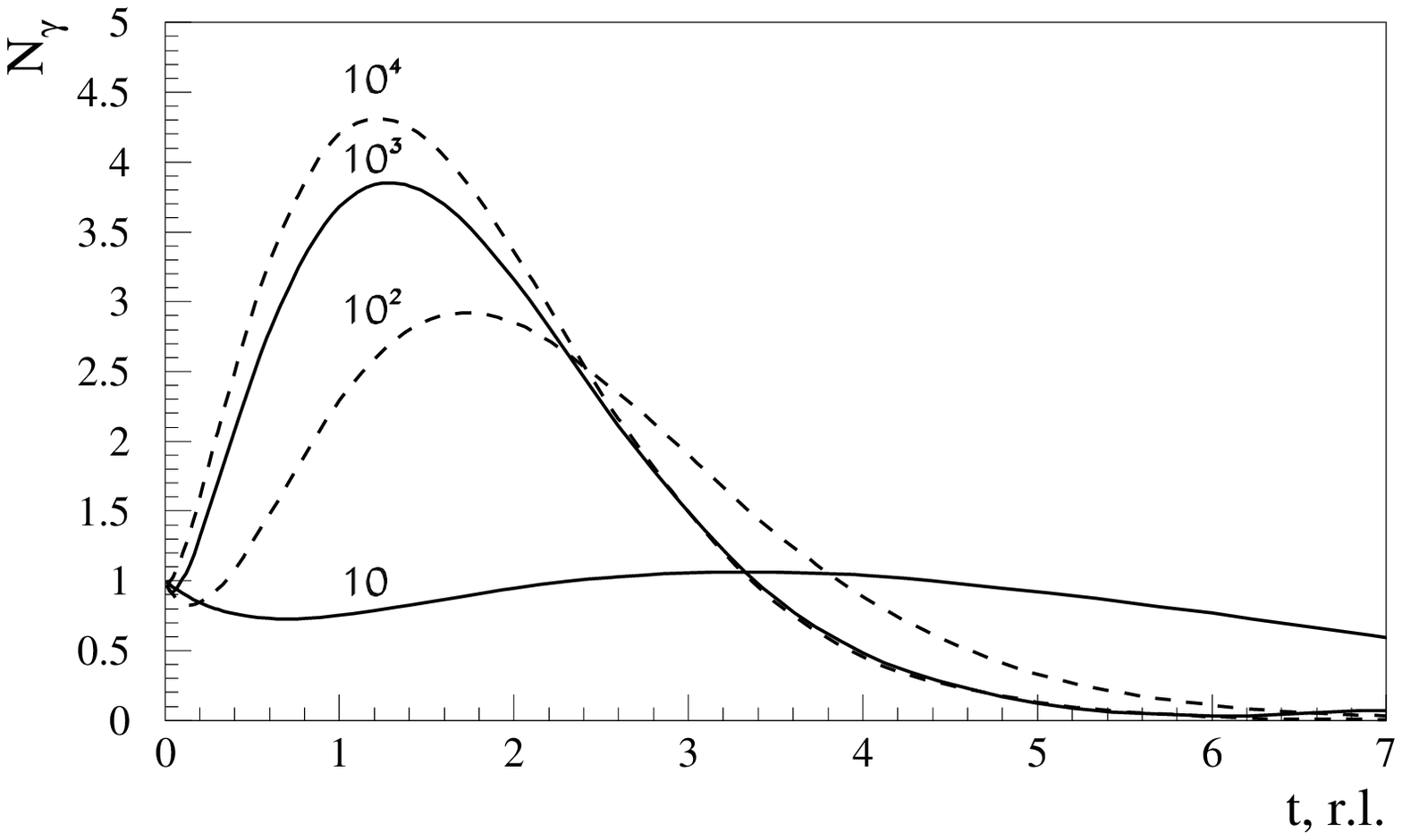} 
\caption{Cascade curves of electrons (upper panel) and
photons (bottom panel) for showers produced in the 
Planckian gas of ambient photons by high energy  photons.  
The results are obtained for 
$\kappa_{th}=\varepsilon_{th}\bar{\omega}_0=1$,
and several  values of  
$\kappa_0=\varepsilon_0\bar{\omega}_0$ indicated at  the curves.
} 
\label{fgr9}
\end{figure}

The results shown  in Fig.~\ref{fgr9}  are cascade curves of electrons 
and photons 
in the blackbody  photon gas calculated for the  fixed value of  
$\kappa_{\rm th}=\varepsilon_{\rm th} \bar{\omega_0}=1$ 
corresponding to the  threshold  energy of   cascade particles,
$ \varepsilon_{\rm th} = m_{\rm e} c^2/3 kT$,  but  for different 
values of the  parameter $\kappa_0$, i.e. for different primary 
photons energies $\varepsilon_0=\kappa_0\cdot m_{\rm e} c^2/3kT$.  

We can see that the  depth $t_{\rm m}$ of the shower maximum 
measured in units of radiation length $X_0^{\rm ( G)}$
depends rather weakly on the primary energy $\varepsilon_0$. It
ranges within $\sim 1\div 2$.  This means that in 
geometrical  units of  length $t_{\rm m}$ is approximately 
a linear function of $\varepsilon_0$. This is explained by  
the  approximately linear  decrease of the cross-sections  
and correspondingly by the  increase of  mean free path of electrons 
and $\gamma$-rays interacting with the ambient photons in the 
Klein-Nishina (quantum) regime (see Fig.~\ref{fgr2}).    

The number of cascade particles in the Klein-Nishina regime 
increases with $\varepsilon_0$ slowly. Even near the shower 
maximum this number does not exceed a few particles.
This is explained by an extremely  high efficiency of conversion 
of energy of the photon to the leading electron at $\gamma$-$\gamma$ 
interactions, and vice versa - the electron energy to the upscattered photon 
at the Compton scattering (see  Figure~\ref{fgr5}  
and Tables 3,4). As a result,  
the energy of the second  (secondary)  particle appears too small
for noticeable contribution to the cascade development. 
Since in  this regime the cross-sections of  the Compton scattering and 
photon-photon pair production  are quite similar, the cascade process 
in the  zeroth approximation can be considered as propagation of a 
single composite $\gamma/e''$  particle which spends 2/5 of its time in the 
``$\gamma$ state'' and 3/5 in the ``electron state'' 
(these times are determined by the 
ratio 1.5 of the corresponding cross-sections in the Klein-Nishina limit).

\begin{figure}[htbp]
\centering
\includegraphics[height=0.30\textheight]{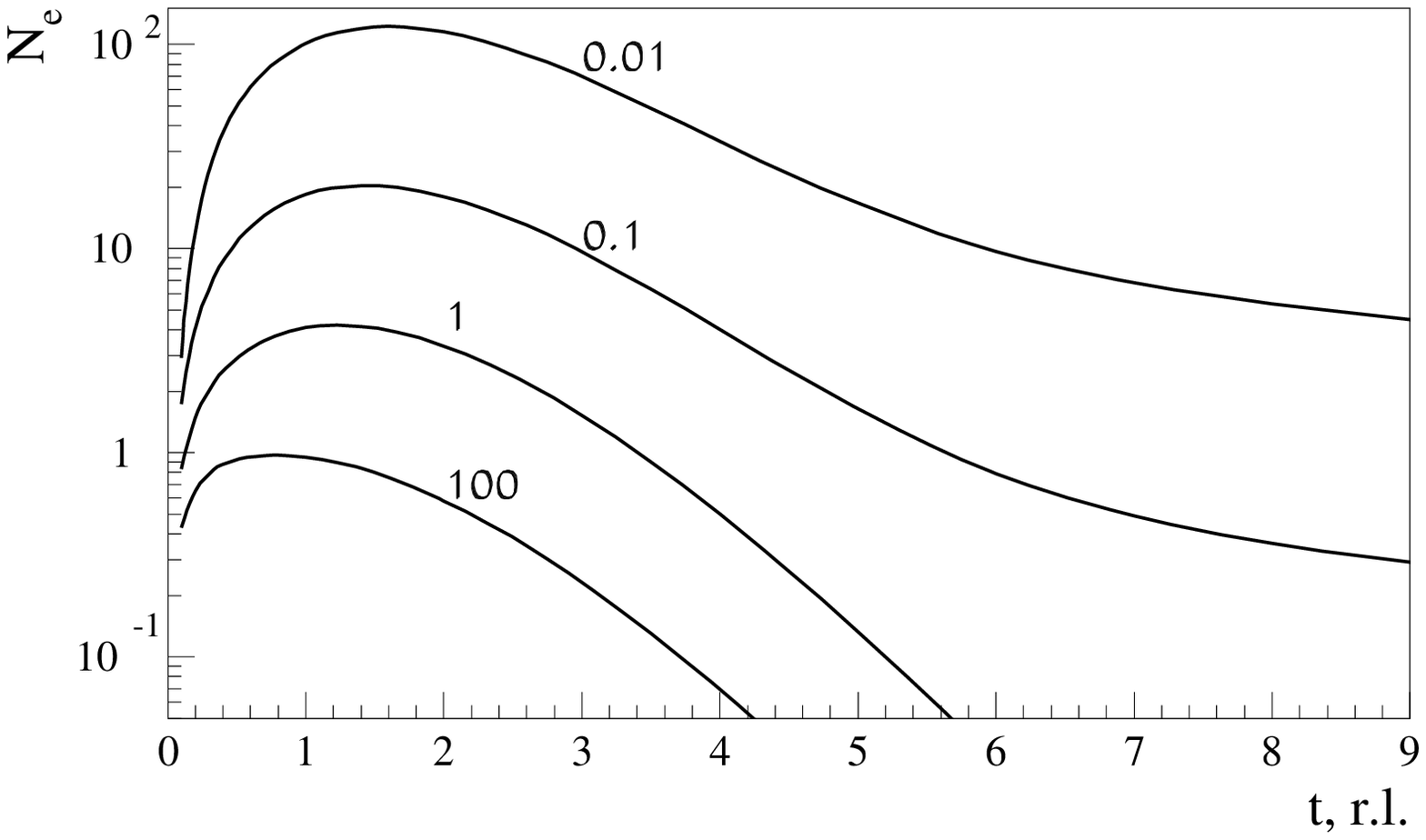} 
\includegraphics[height=0.30\textheight]{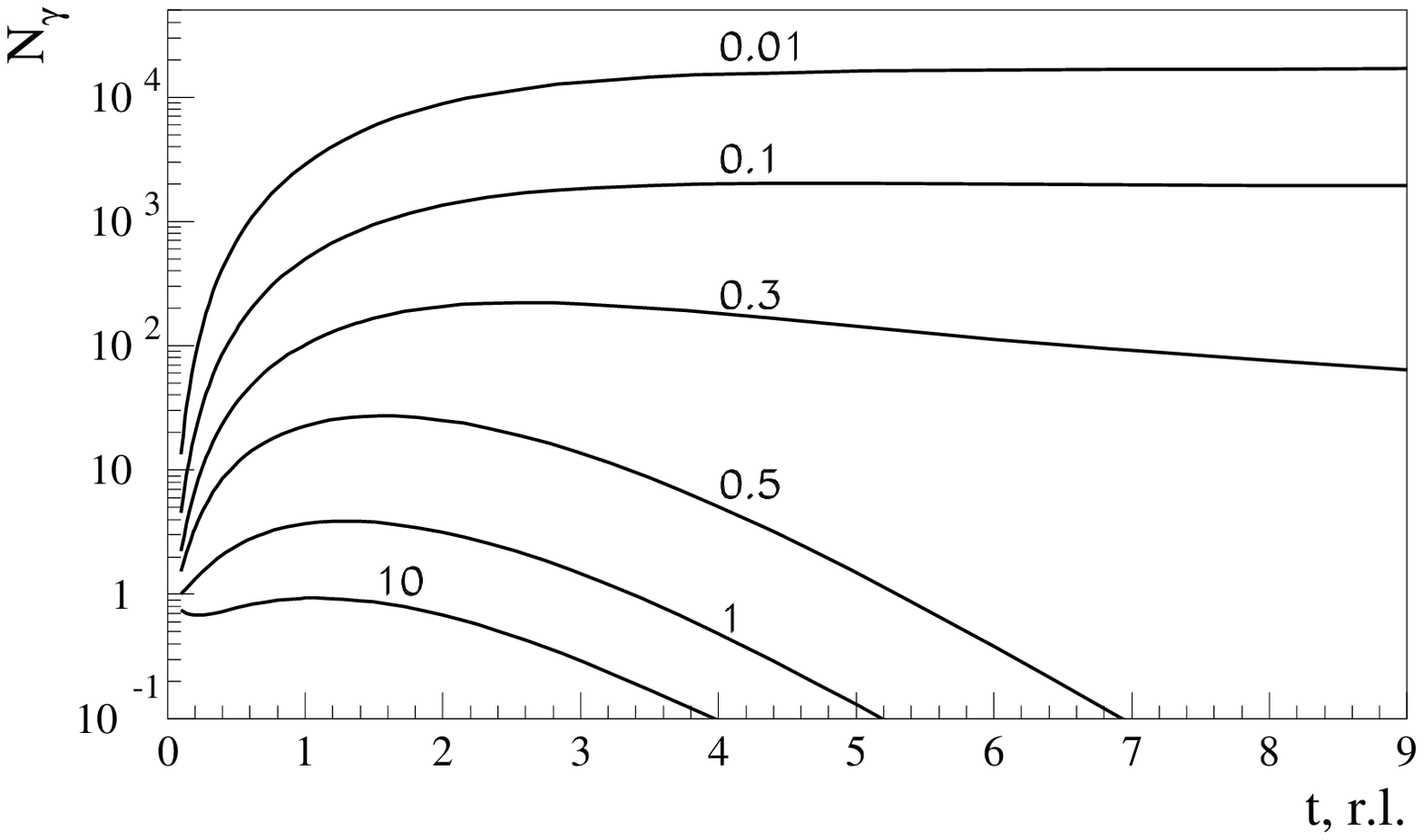}  
\caption{Cascade curves of electrons (upper panel)
and photons (bottom panel) for  showers produced in 
the Planckian  gas of ambient photons by  high energy 
photons.   The results are obtained for 
$\kappa_0=\varepsilon_0\bar{\omega}_0=10^3$ and for 
values of 
$\kappa_{th}=\varepsilon_{th}\bar{\omega}_0$ indicated at 
the  curves.}
\label{fgr10}
\end{figure}

Fig.~\ref{fgr10} illustrates dependence of the cascade 
curve on the threshold energy $\varepsilon_{\rm th}$
with the corresponding parameter $\kappa_{\rm th}$
below and above 1.   With reduction of  $\kappa_{\rm th}$, 
the cascade curves of both $\gamma$-rays and electrons increase,
especially in the regime of   $\kappa_{\rm th} \leq 1$. This has a simple 
explanation.  Although in this (Thompson) regime of scattering 
the cross-section is large,   the energy 
transferred to the secondary photon is quite small (see Table~3). 
Consequently,  the mean free path of these electrons  increases with 
reduction of their energy as $1/\varepsilon$. These electrons produce 
large number of $\gamma$-rays with  energy below the {\em effective}
pair-production threshold \footnote{In the case of the black-body radiation  
of background  photons there is no strict kinematic threshold;  
$\gamma$-rays  with energy $\varepsilon \leq m_{\rm e}c^2/3 kT$
may still interact with the background photons from the Wien tail 
of distribution.}, which therefore penetrate very large distances
without interacting with the ambient background radiation. 
Consequently, the number of these photons remains almost constant.

%% Figure 9
%% ----------------------------------------------------------------------------

% Figure 10.
% ----------------------------------------------------------------------------

\subsubsection{Magnetic field}

\begin{figure}[htbp]
\centering
\includegraphics[width=0.65\textwidth]{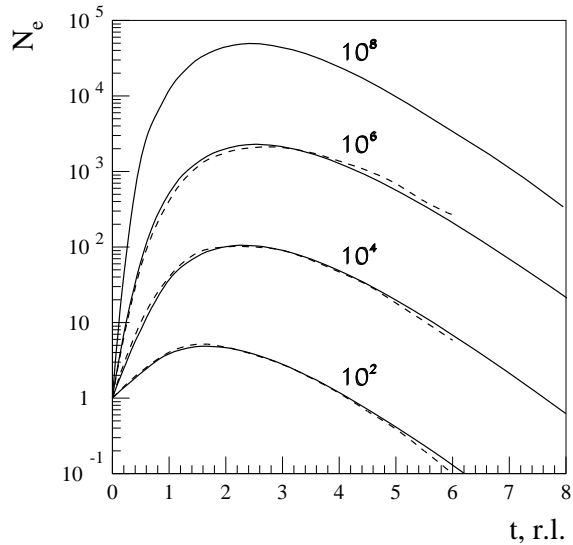} 
\caption{Cascade curves of electrons for  showers initiated by
primary electrons in the magnetic field. Different values 
(indicated at the curves) of the ratio of primary and 
threshold energies are assumed  for the fixed  
$\chi_{th}=\varepsilon_{th} H/H_{\rm cr}=10^3$.
For comparison, the results obtained in Ref. \cite{Anguelov}
are also shown  (dashed curves).
}
\label{fgr11}
\end{figure}

Fig.~\ref{fgr11}  shows the cascade  curves  of electrons in the  magnetic field
obtained with  the adjoint equation technique.  The  comparison with the 
Monte Carlo simulations \cite{Anguelov} shows nice agreement
between  results  obtained by  two different methods.   

Following to Ref. \cite{Anguelov} we express 
the depth of penetration of cascade particles in units of the     
radiation length in the magnetic field
\begin{equation}
X_0^{\rm (F)}=0.207\cdot 10^{-7}\frac{H_{\rm cr}}{H}\chi_0^{1/3} \ ,
\label{rlB}
\end{equation} 
where $\chi_0=\varepsilon_0 H/H_{\rm cr}$.
Similar to the cascade in the photon gas,  the radiation length
in the magnetic field depends on the primary energy, but in this
case the dependence is slower, $X_0^{\rm (F)} \sim \varepsilon_0^{1/3}$.

% Figure 11.
% ----------------------------------------------------------------------------

Fig. \ref{fgr11} illustrates dependence of cascade curves 
on the primary energy in a deep quantum regime 
with the threshold  value of the parameter  $\chi_{\rm th}=10^3$.  
It is seen  that the location  of the cascade curve maximum  $t_{\rm m}$
ranges within $2\div3$ radiation lengths. In geometrical units this implies  
that  the position of the cascade curve maximum 
increases with  energy proportional to $\varepsilon_0^{1/3}$. 
The maximum electron number, $N_{\rm m}$,  grows with $\varepsilon_0$ 
as  $\sim\varepsilon_0^{0.7}$. 
Thus,  the cascade development in the magnetic field in the quantum regime  
is somewhat intermediate between the cascades in  matter and the photon gas.

% Figure 12.
% ----------------------------------------------------------------------------
\begin{figure}[htbp]
\centering
\includegraphics[height=0.30\textheight]{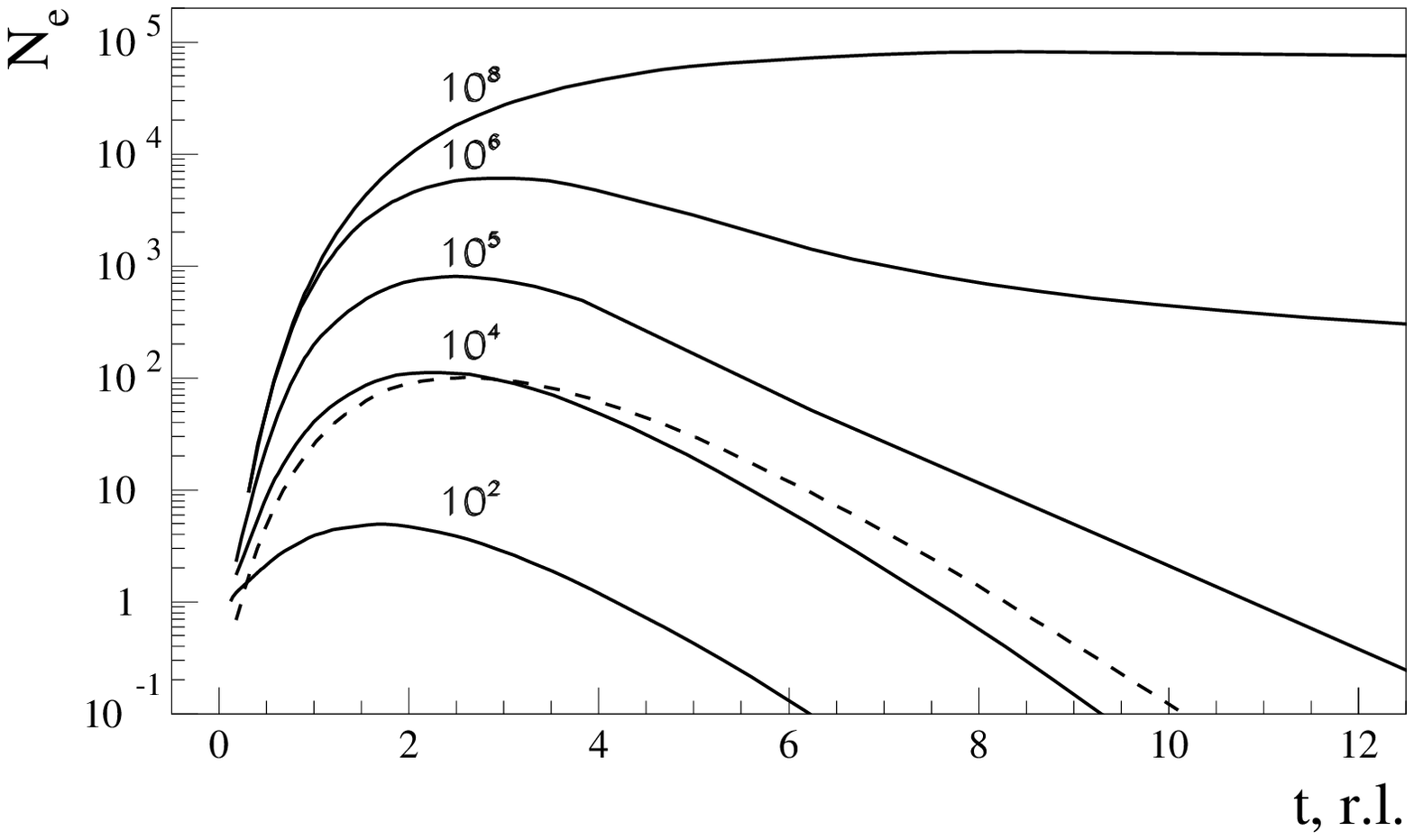} 
\includegraphics[height=0.30\textheight]{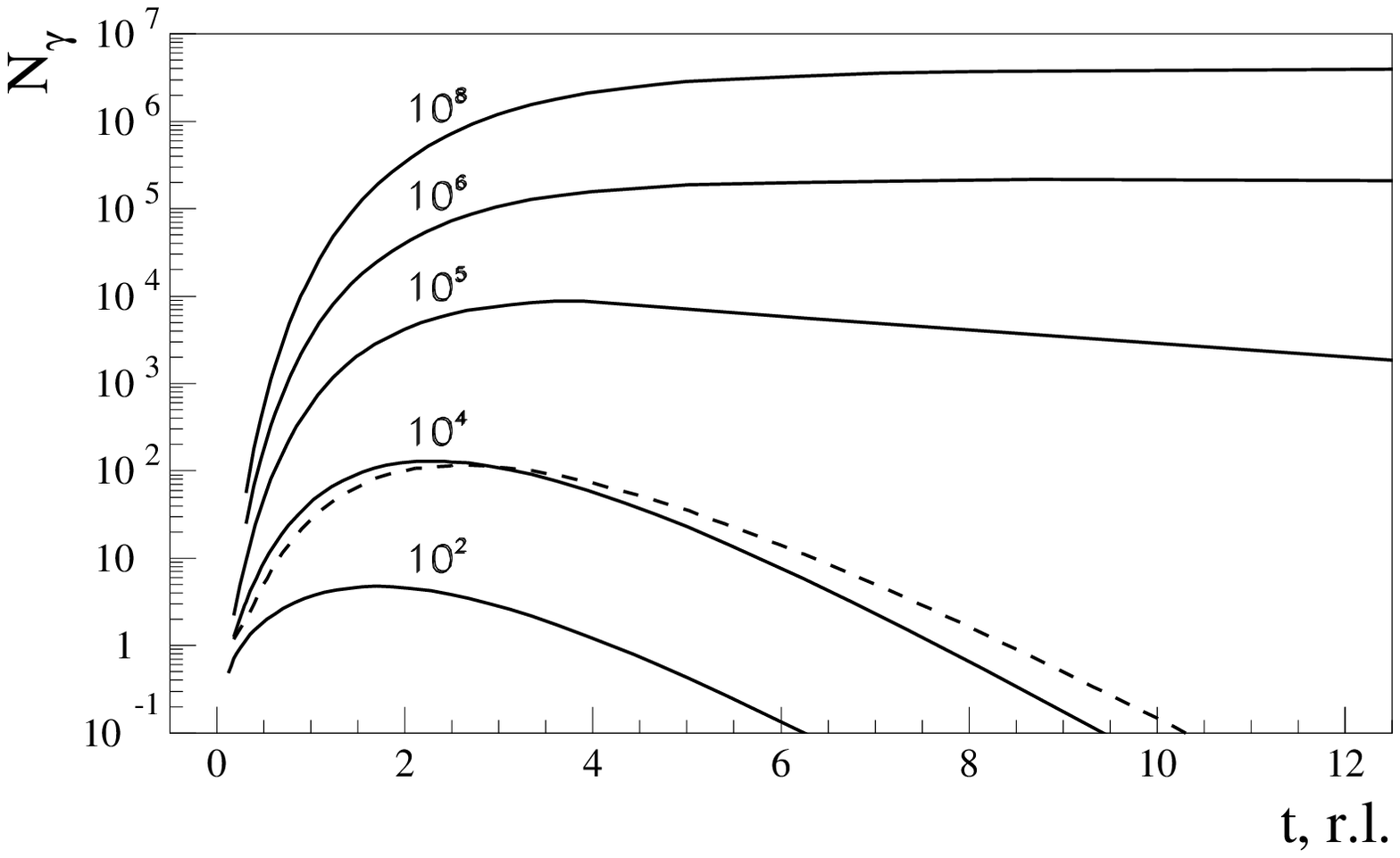}  
\caption{Cascade curves of electrons (upper panel) 
and photons (bottom panel)   for showers  initiated by
primary electrons (solid curves) 
and photons (dashed curves) in the magnetic field. Different values
of the threshold energy ($\varepsilon_{\rm th}$) are assumed for the 
fixed $\varepsilon_0=2\cdot 10^{7}$ and  $H=10^{11}$~G.
The   $\varepsilon_0/\varepsilon_{\rm th}$ ratios are  indicated at the
curves. The corresponding values for  $\chi_{\rm th}$ are presented 
in Table~5.} 
\label{fgr12}
\end{figure}

% Table 5. 
% -----------------------------------------------------------
\begin{table}
\caption{The ratio $N_{\gamma}/N_e$  for cascade curves 
presented in Figure~12.}
\begin{center}
\begin{tabular}{lccccc} \hline 
$\varepsilon_0/\varepsilon_{\rm th}$& 
$10^2$         & $10^4$&   $10^5$    &    $10^6$&  $10^8$   \\ \hline
$\chi_{\rm th}$& $4.5\cdot 10^2$  &4.5    &  
$4.5\cdot 10^{-1}$&$4.5\cdot 10^{-2}$&$4.5\cdot 10^{-4}$ \\ \hline
$t=$0.3 r.l. &  0.76& 0.92& 1.44& 2.66& 5.95  \\
$t=$0.9 r.l. &  0.94& 1.09& 2.73& 4.65& 1.92  \\
$t=$3 r.l. &  1.00& 1.17& 11.2& 18.6& 45.2  \\
$t=$6 r.l. &  1.03& 1.19& 107 & 140 & 455  \\ \hline
\end{tabular}
\end{center}
\end{table}

In Fig.~\ref{fgr12}  we present the cascade curves of electrons and photons 
for different values of the threshold energy 
$\varepsilon_{\rm th}$.
It is seen that in the regime  $\chi_{\rm th} \ll 1$ the  cascade 
curves  have a behaviour quite similar to that for the cascade curves
in the photon gas  shown in Figure~10. This can be  explained  by similarities 
of the synchrotron radiation and the Compton scattering 
in the non-quantum regime.  In both cases the electrons
suffer a large number of collisions but with small energy transfer
to secondary photons. In addition, in both cases  $\gamma$-rays 
stop to interact  effectively  with the ambient medium.
  
Table~5  demonstrates the dependence of the 
$N_{\gamma}/N_{\rm e}$ ratio  on $\chi_{\rm th}$ and 
$t$.  It behaves quite similar to cascades in  matter. 
Particularly, for large threshold energies $\varepsilon_{\rm th}$ 
the electron and  photon numbers  are  comparable; 
for small $\varepsilon_{\rm th}$ the photon number is considerably
larger, and  the ratio $N_{\gamma}/N_{\rm e}$ increases with the depth.

\subsection{Energy spectra of cascade particles}

Compared to the cascade curves, the energy spectra 
of electrons and photons  at different 
stages  of the cascade  propagation contain more
circumstantial   information  about the
shower characteristics.    

 % Figure 13.
% -----------------------------------------
\begin{figure}[htbp]
\centering
\includegraphics[width=0.65\textwidth]{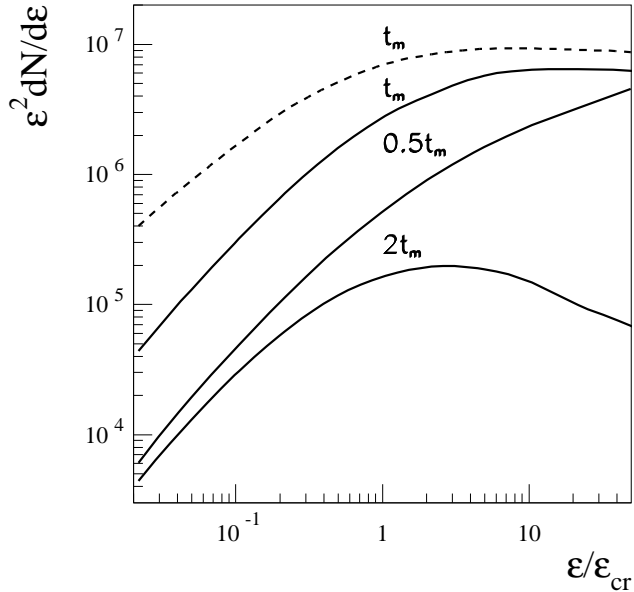} 
\caption{Differential energy spectra of 
electrons (solid curves) 
and photons (dashed curve) for cascades initiated in 
the hydrogen gas  by
a primary photon  with energy $\varepsilon_0=2\cdot 10^8$. 
Different values of depths in units of the cascade maximum $t_{\rm m}$ 
are shown at the curves.}
\label{fgr13}
\end{figure}

In Fig.~\ref{fgr13}  we show  the  energy spectra of 
cascade electrons and photons at different 
penetration depths in  the hydrogen gas.  At energies 
exceeding the critical energy,   $\varepsilon \gg \varepsilon_{\rm cr}$,  
the spectra of both electrons and photons 
are described by power-law  $dN/d\varepsilon\sim \varepsilon^{-\alpha}$,
where  the spectral index is function of the penetration depth. 
Near the shower maximum,  $\alpha\simeq $2 for both electron and 
photon spectra. With depth the spectra become  steeper.

% Figure 14.
% ----------------------------------------------------------------------------
\begin{figure}[htbp]
\centering
\includegraphics[height=0.3\textheight]{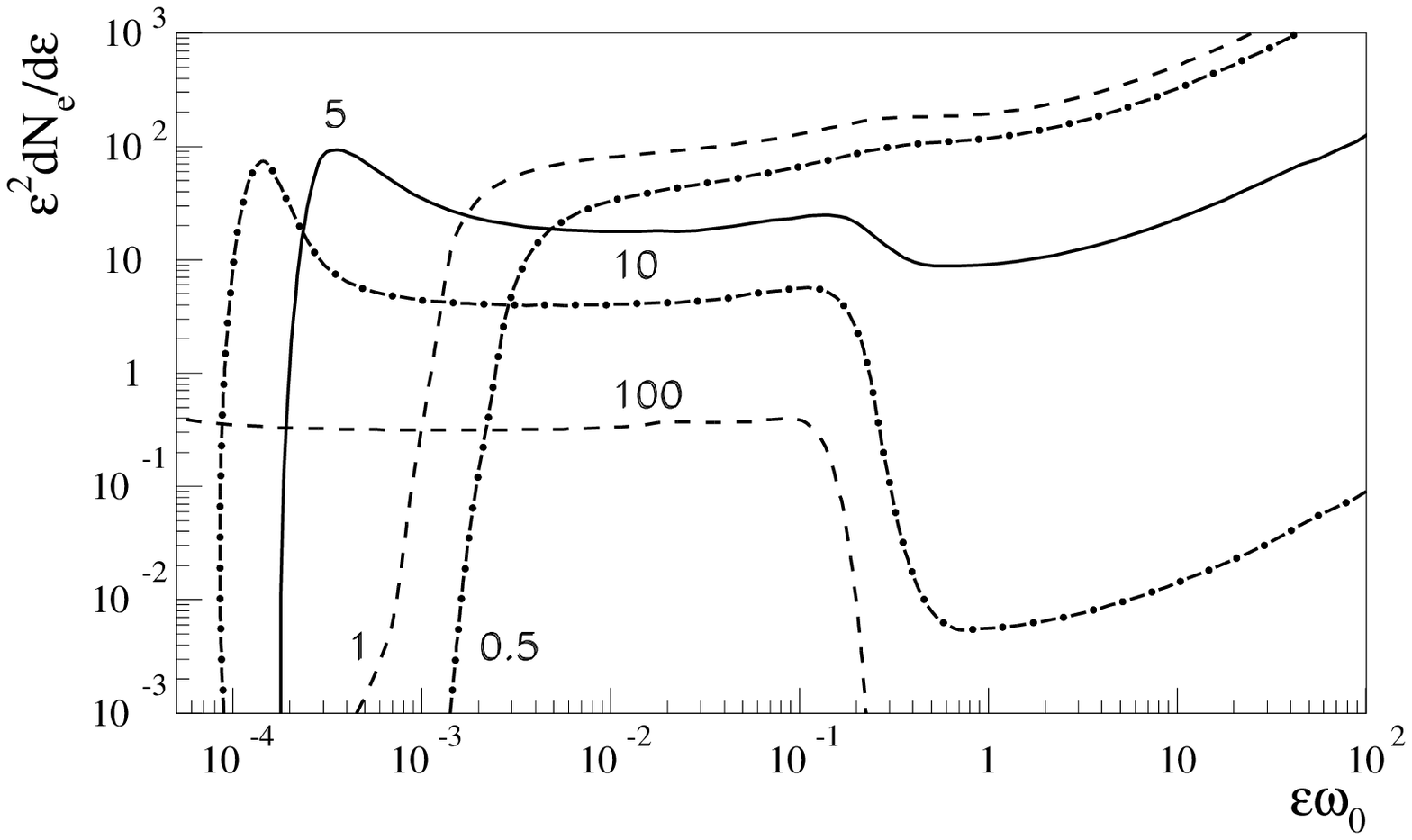} 
\includegraphics[height=0.3\textheight]{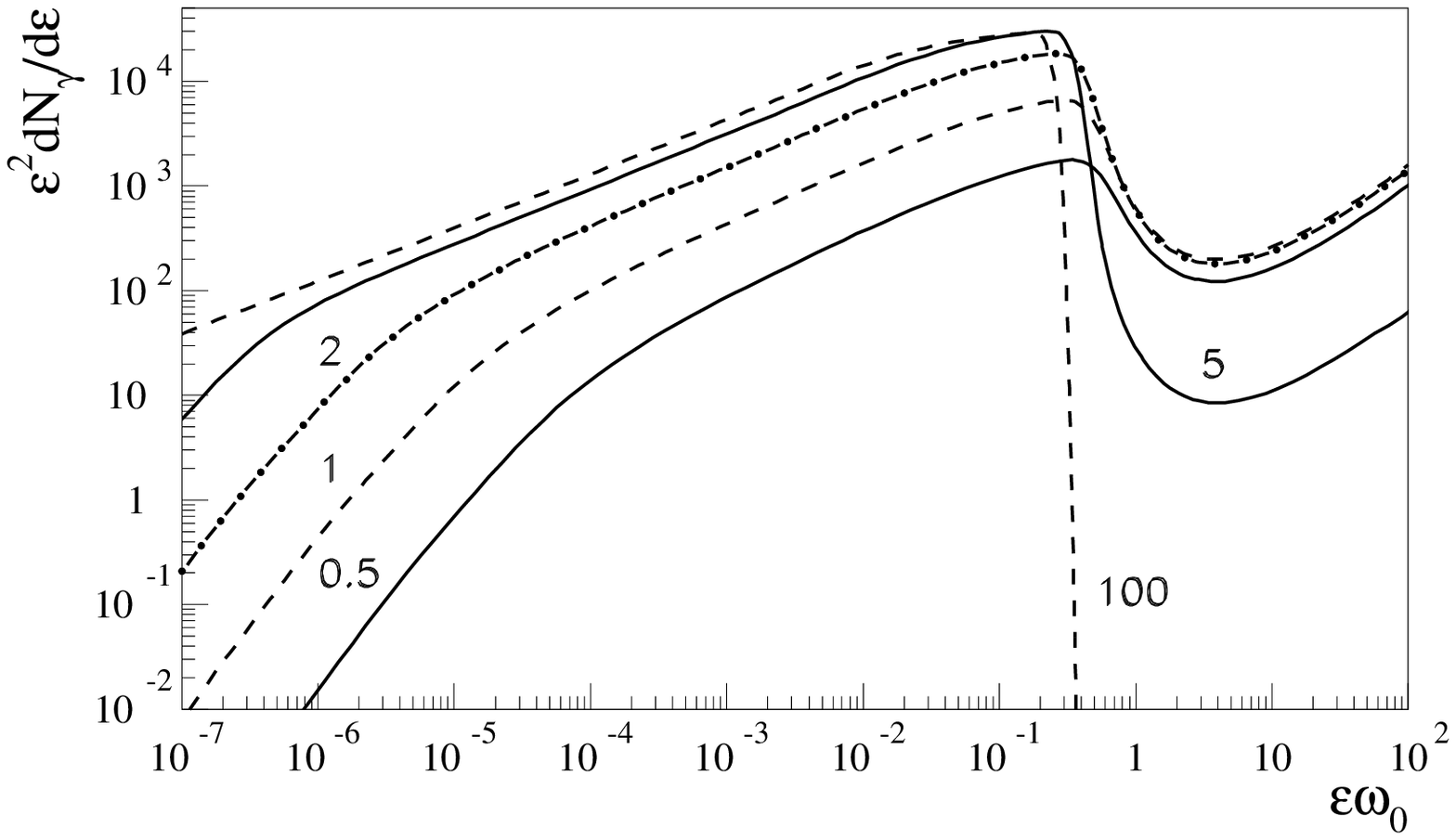} 
\caption{Differential energy spectra of cascade electrons 
(upper panel) and photons (bottom panel)    in the  cascade initiated   
by a primary photon ($\kappa_0=10^3$)
in the  radiation field with Planckian spectral 
distribution. The spectra are calculated
for different penetration depths indicated (in units of radiation lengths) 
at the curves.}
\label{fgr14}
\end{figure}

Below the critical energy,   the cross-sections of the  
bremsstrahlung and pair-production processes are  
not  sufficiently large  to support the cascade development 
against the dissipative processes  like  ionization and  Compton  scattering. 
Thus the  multiplication of  cascade particles is dramatically reduced.  The 
cooling time of electrons due to ionization losses is proportional to 
energy,  therefore below the critical energy this process leads to   
significant hardening of  the electron, and correspondingly 
also the photon spectrum that behaves as 
${\rm d}N/{\rm d}\varepsilon \propto  \varepsilon^{-1}$.

In Fig.~\ref{fgr14} we  show  the differential energy spectra of 
cascade particles in the radiation field with Planckian type 
spectral  distribution.    These  spectra are quite 
different from the ones that appear  in  the cascades 
developed in matter. 
In the high energy (Klein-Nishina) region,  
$\kappa=\varepsilon \omega_0 \gg 1$, 
and for not very large depths ($t\le 10$~r.l.) the shape of  
the differential spectra of electrons and 
photons is quite insensitive to the depth (contrary to the 
cascade in matter).   At these energies  the 
differential spectra  are   characterized by slopes with
indices  between 1 and 1.5.

At $\kappa\sim 1$  there exists a pronounced  transition region where
the energy spectra undergo dramatic changes. The reason for 
the appearance of the transition region is  the change of the character of 
Compton scattering (from the Thompson to the Klein-Nishina regime)
and the  sharp  reduction of  the pair production cross-section. 
Obviously,  a broader energy distribution of background photons 
would make this transition smoother and less pronounced.  

% Figure 15.
% ---------------------------------
\begin{figure}[htbp]
\centering
\includegraphics[height=0.3\textheight]{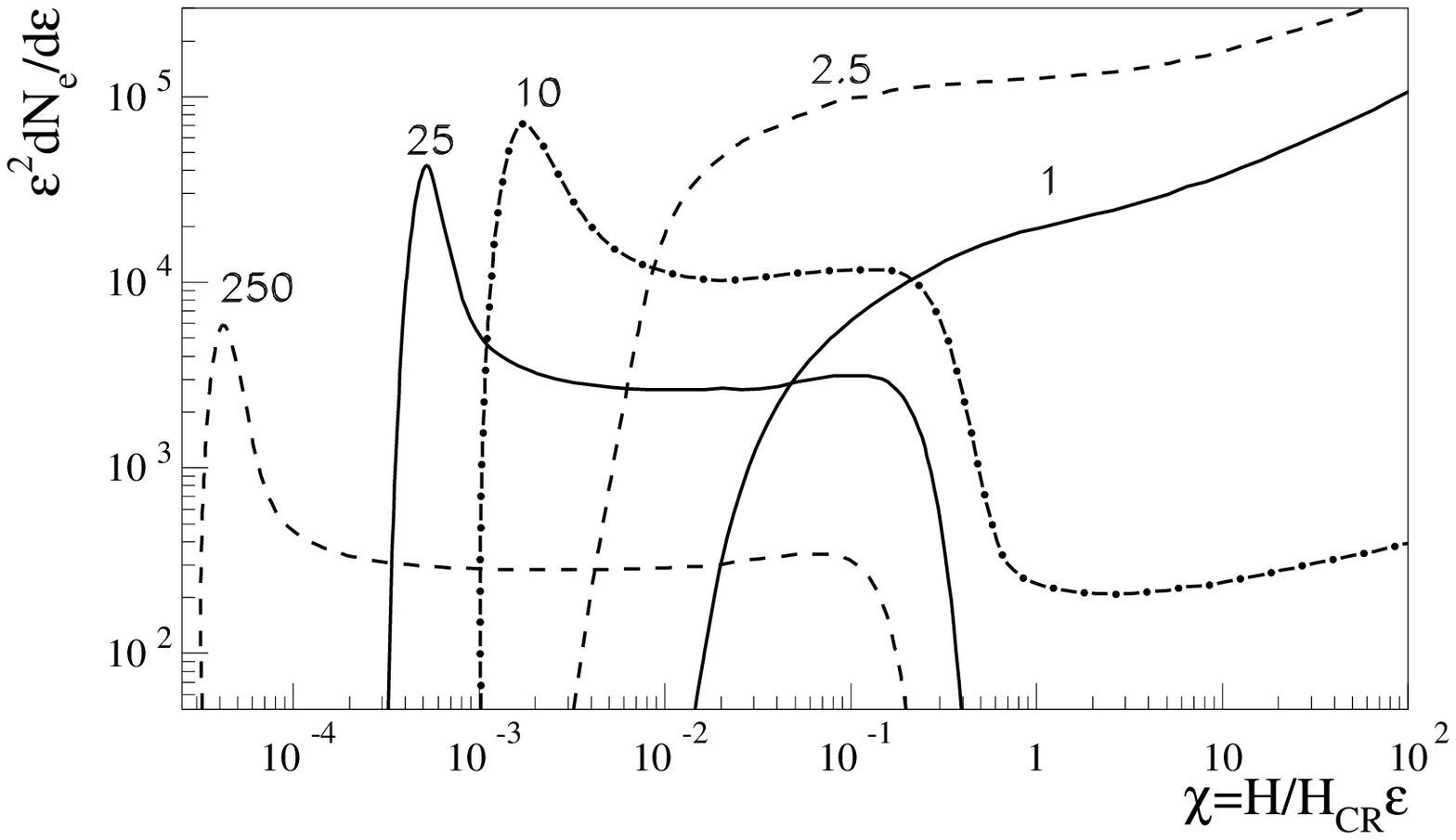}
\includegraphics[height=0.3\textheight]{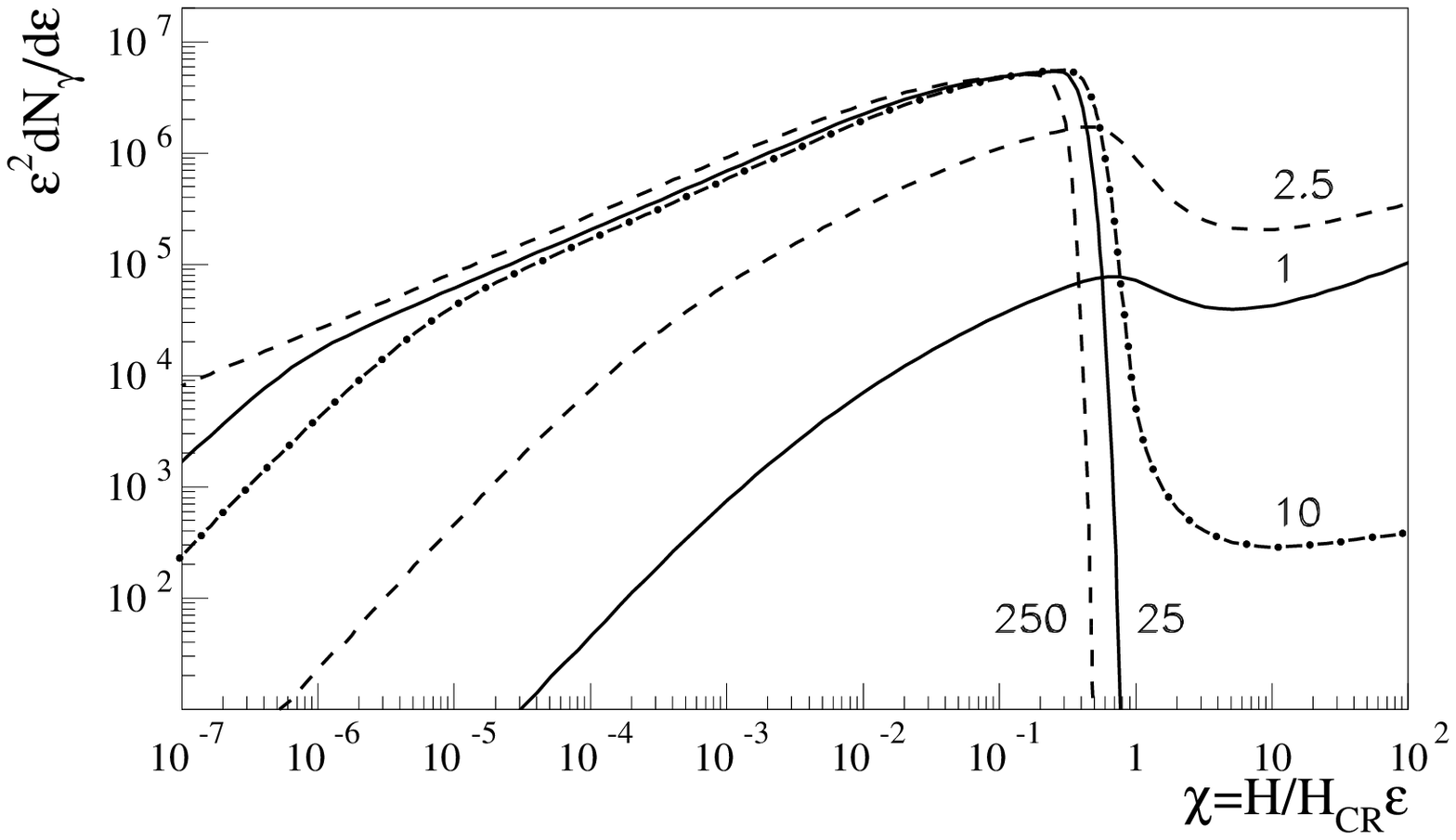} 
\caption{Differential energy spectra of cascade 
electrons (upper panel)  and photons (bottom panel)   in showers  
initiated  in the  magnetic field of intensity $H=10^{11}$~G by
a primary photon  of energy $\varepsilon_0=2\cdot 10^7$. 
The spectra are calculated
at different depths indicated (in units of radiation lengths) 
at the curves.}
\label{fgr15}
\end{figure}

At low energies   ($\kappa\sim 1$) the cascade 
development is not supported by pair-production.
Here  we deal with  an ensemble of photons 
not interacting with the  environment and  an ensemble of electrons 
continuously cooling  down with the characteristic (Thompson) 
time $t_{\rm T} \sim 1/\varepsilon$.  This results in    standard 
electron  spectra  with spectral index $\alpha=2$. 
During  propagating into deeper layers of the photon gas,  
these electrons  produce  large number  
of low  energy photons  with energy spectrum  
${\rm d} N_\gamma/{\rm d} \varepsilon \propto \varepsilon^{-1.5}$.

For any  finite depth $t$, electrons do not have enough time to be cooled 
down to energies $\varepsilon \to  0$. Therefore  the electron spectrum 
drops dramatically below  some energy $\tilde{\varepsilon}_{\rm e}$ 
which  decreases  with depth  as $\sim 1/t$. This effect is clearly seen 
Fig.~\ref{fgr14}. The corresponding response in the 
photon spectrum is also  quite distinct. Below the break energy around 
$\sim \omega_0  \tilde{\varepsilon}_{\rm e}^2$, the 
photon spectrum becomes extremely hard, 
${\rm d} N_\gamma/{\rm d} \varepsilon \approx const$.

The energy spectra of electrons and photons produced during 
the cascade development in the magnetic field  are  quite similar
to the spectra  of electromagnetic 
cascades in the  radiation field   with Planckian distribution of target photons.
These spectra are shown in   Fig. \ref{fgr15}.  All features discussed above 
in the context of  cascading  in the photon gas,  are clearly seen also 
in cascades developed in the magnetic field, if we 
express the penetration depths in units of radiation lengths, and the  
energies of electrons and photons in the form of products 
$\varepsilon \omega_0$ and $\varepsilon (H/H_{\rm cr})$. The cascade  
spectra in radiation and magnetic fields are not, however, identical.
For example, because of significant  differences in    
the asymptotics  of relevant cross-sections, 
\gr  spectra in the magnetic field in the quantum regime 
are flatter   than the corresponding \gr  spectra in the photon gas 
in the Klein-Nishina regime (compare Figs. \ref{fgr14} and \ref{fgr15}).

\section{Mixed  environment}

In order to reveal  peculiarities of the cascade development 
in different substances, in previous sections we limited our 
discussion to  the ``clean''  environments dominated by matter,  
radiation or magnetic fields.   The  ``pure cascade'' concept is   
not  only a convenient  theoretical  approximation. In fact,  under 
certain realistic conditions and within limited energy regions,  
this could be  the most likely realization of particle interactions
with then ambient medium.   The relativistic electron-photon cascades in 
the Earth's atmosphere,  in the intergalactic medium   and in 
pulsar magnetospheres are 3 characteristic examples of ``pure'' 
cascade  developments in matter, photon gas and magnetic field, 
respectively.

In some  cases, however,  ``parallel'' interactions  of electrons 
and   photons with  2 or 3 substances can proceed simultaneously 
and with  comparable  efficiencies.
The outcome of the  interference of several competing processes could 
be  quite different and complex  depending on relative  densities 
of the ambient plasma, radiation and  magnetic fields,  as well as  on the 
energy  of primary particles.
For example, interactions of $\geq 10^{20} \ \rm eV$ protons with 2.7 K CMBR 
leads to production of secondary  electrons, positrons and $\gamma$-rays of 
energy $\geq  10^{19} \ \rm eV$,  which in their turn 
trigger electromagnetic cascades in the same radiation field. However,
due to the synchrotron  cooling of electrons this process  would be 
significantly suppressed if the intergalactic magnetic  field exceeds 
$10^{-10} \ \rm G$. The characteristic energy of   synchrotron photons  
$\sim 5 \times 10^{8} (B/10^{-10} \ \rm G) (E/10^{19} \ \rm eV)^2 \ \rm eV$
is too small for interactions with the diffuse 
photon fields on the Hubble scales.   
However, if the energy of electrons and/or photons 
injected into the intergalactic medium exceeds $\geq 10^{21} \ \rm eV$
(this could be the case, for example,  of secondary products  from decays of the
so-called  topological defects), then the synchrotron photons 
appear  in the TeV energy range. The TeV \grs  effectively interact with the diffuse 
extragalactic  infrared radiation, and trigger  new, low energy electron-photon 
cascades. In both cases the synchrotron radiation changes the character 
of the cascade development in the radiation field, but cannot support 
its ``own'' cascade  in the magnetic field.  

Below we briefly discuss a more 
interesting  scenario, 
when the cascade  develops both in the radiation and magnetic fields.    
Let's assume that a $\gamma$-ray photon of energy 
$E_{\gamma}= 10^{20}$~eV    is injected into a highly magnetized 
low-density plasma with thermal radiation density comparable to the 
energy density of the magnetic field $H^2/8 \pi$. In principle such a 
situation  can occur  in the vicinity  of central  engines of AGN.

% Figure 16.
% ----------------------------------------------------------------------------
\begin{figure}[htbp]
\centering
\includegraphics[height=0.3\textheight]{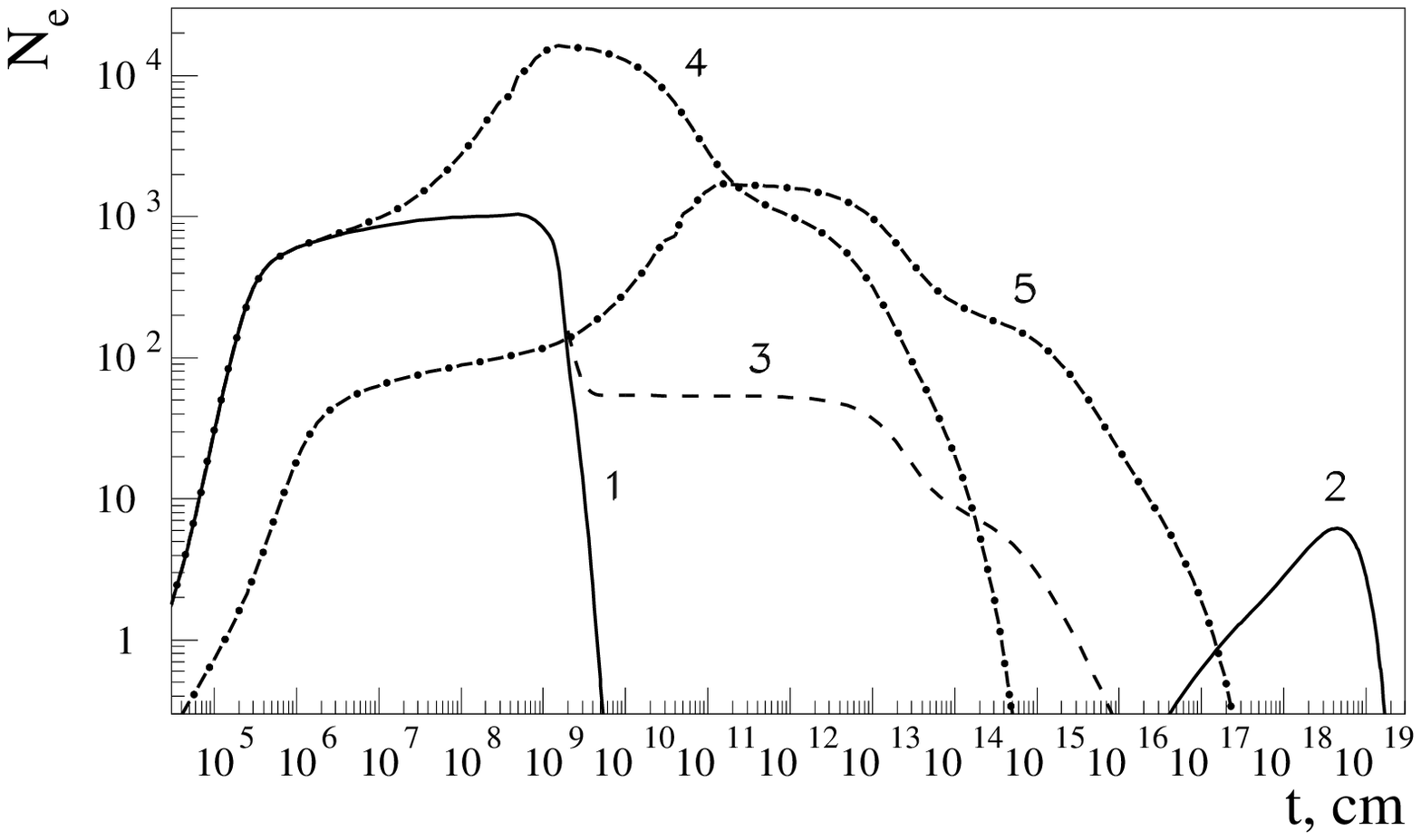} 
\includegraphics[height=0.3\textheight]{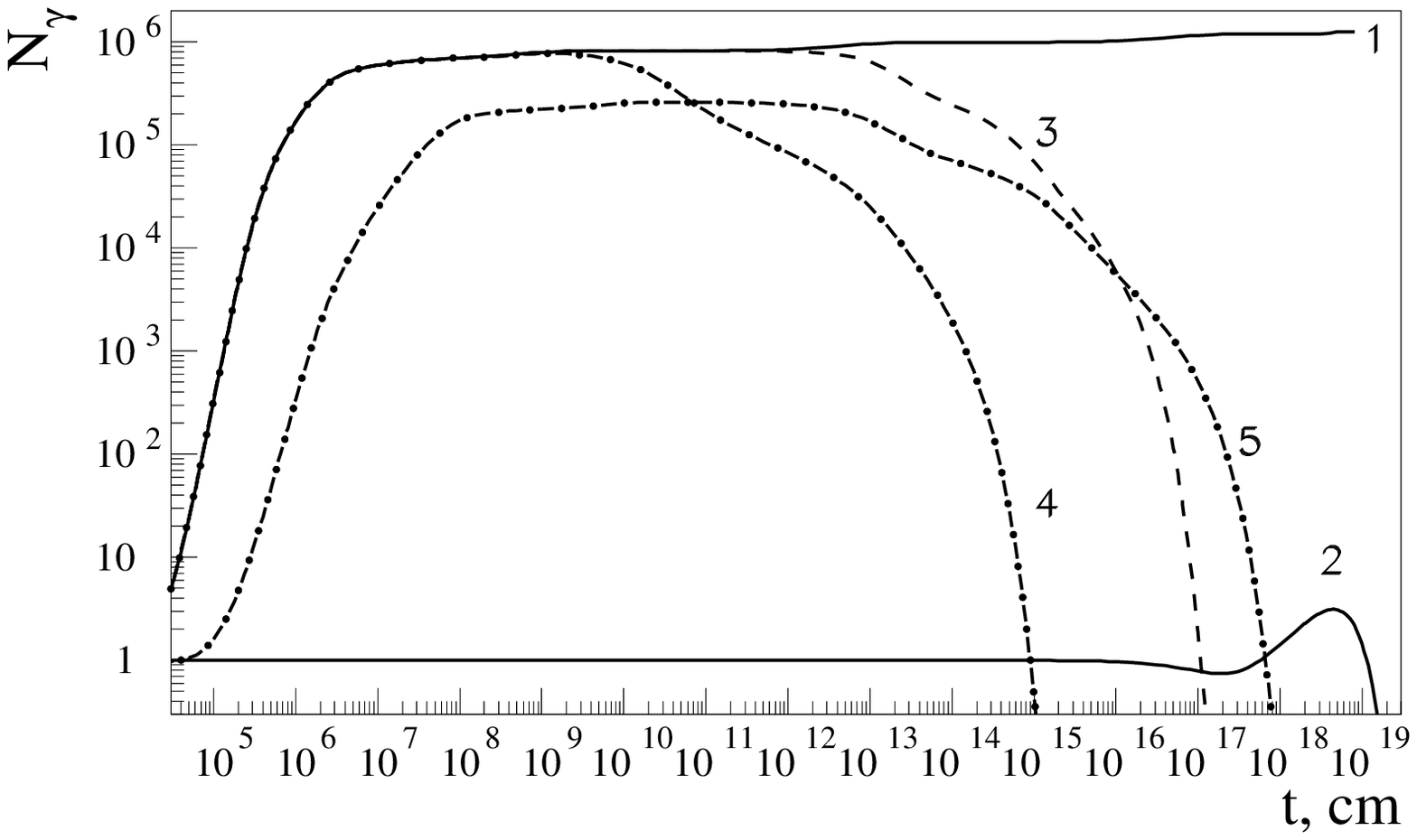} 
\caption{ The cascade curves of electrons (upper panel) and 
photons (bottom panel) with energy $E\ge E_{\rm th}=5\cdot 10^{11}$~eV
for showers  initiated by a primary photon of energy 
$E_0=10^{20}$~eV  in a mixed environment 
consisting of the magnetic field 
and  the blackbody radiation. The calculations are performed for 
5 combinations of parameters characterizing the  target 
radiation and magnetic fields. These parameters   are listed  
in Table~6.} 
\label{fgr16}
\end{figure}

% Table 6. 
% -----------------------------------------------------------
\begin{table}
\caption{Parameters of the mixed environment used for calculations 
of the cascade curves shown  in Fig.~\ref{fgr16}. $H$ is the strength  of 
the magnetic field, $T$ is the temperature of the black-body radiation, 
$w_{\rm F}$ and  $w_{\rm G}$ are energy densities of the magnetic field 
and black-body  radiation, respectively.}
\bigskip
\begin{center}
\begin{tabular}{lccccc} 
\hline
Parameter Combination & 1  &  2&        3&   4&   5   \\ \hline
$H$, Gauss            & 100&       0&  100& 100& 10 \\ 
$kT$, eV    &   -- &       3&    0.3&   3& 0.3  
\\ 
$w_F$, erg/cm$^3$ &  400& 0 &    400&  400&   4 \\ 
$w_G$, erg/cm$^3$ &   0& $\simeq 10^4$&$\simeq  1$ &
$\simeq 10^4$& $\simeq 1$ \\ \hline
 
\end{tabular}
\end{center}
\end{table}

Fig.~\ref{fgr16}  illustrates the impact of 
the magnetic field and the temperature 
of blackbody  radiation on cascade curves. 
Five different  combinations of these parameters 
presented in Table 6  have been analyzed.  
The minimum particle  energy  for the  
cascade  curves was  taken $5 \times  10^{11}$~eV, i.e.
comparable or larger (for the assumed radiation temperatures)  
to the  effective pair  production  threshold in the  
black-body radiation ,  
$E_{\rm th}^{\rm (G)}  \sim  m_{\rm e }^2c^4/kT \simeq  
10^{11} (kT/1 \ \rm eV)^{-1} \ \rm eV$. 

For the chosen parameters,  
the radiation length in  the magnetic field is much  
smaller than the mean free path  
of primary $\gamma$-rays  in the photon  
field. At  absence  of magnetic field 
({\it parameter combination~2})  primary \grs   
penetrates  very deep  into the source 
without interacting with  the ambient photon gas.   
Consequently,  the cascade starts very  late and remains as 
underdeveloped.   

The presence  of a strong magnetic field  makes the 
cascade development much more effective, which now is 
supported  by magnetic pair production and synchrotron radiation. 
The cascade develops in this regime  until the energy of $\gamma$-rays
is reduced down to the   effective threshold  of the magnetic pair production,   
$E_{\rm th}^{\rm (F)}\simeq 10^{17} (H/100 \ \rm G)^{-1} \ \rm eV$. 
Therefore, in the case of pure magnetic field we have quite simple 
and predictable cascade curves  ({\it parameter combination~1}). 

The presence of  photon gas changes significantly  the character of 
cascade development.  At energies  
above $E_{\rm th}^{(F)}$ both electrons and \grs interact mainly with 
magnetic field. When the cascade particles enter
the energy interval determined  by the 
pair-production thresholds  in the radiation  and magnetic fields,  
[$E_{\rm th}^{(G)}$, $E_{\rm th}^{(F)}$],  the \grs  
interact effectively with the photon gas, while  
the electrons  continue to interact mainly with the magnetic field. 
Although for  $kT=3 \ \rm eV$,  the blackbody 
radiation density considerably exceeds the 
energy density  of 100 G magnetic field (see Table 6), because
of  the Klein-Nishina effect the interactions of electrons with 
photon gas become significant only when the electrons are cooled 
down to energies $\leq 100$ GeV.       
The mean free path of \grs  in the black-body radiation
field has  minimum around 
$E_{\rm th}^{\rm (G)}$.  Therefore, \grs start to interact intensively with 
the ambient photons at  depths comparable with the mean free path 
 $\Lambda_{\rm min} \sim \Lambda_{\gamma}(E_{\rm th}^{\rm (G)})$.  
This  results in a  rapid 
growth of  the electron cascade curve  and   reduction of the 
high energy photon  cascade curve ({\it parameter combinations~3-5}).
Due to the rapid synchrotron cooling, the mean free paths of 
very energetic electrons in the magnetic field 
are very short, therefore for each given depth the main 
source of these electrons  are  \grs which  
can penetrate much deeper. On the other hand, due to the 
same synchrotron losses, the electrons cannot support 
reproduction of very high energy  $\gamma$-rays. 
Therefore, the at depths $\gg  \Lambda_{\rm min}$
both the electron and photon curves are suppressed.

The  increase of the temperature increases  the  
energy density of the photon  gas  ($\propto  T^4$)  and makes 
interactions  of cascade particles with radiation more intensive. 
This results in  faster absorption of 
cascade particles ({\it parameter combination~4}).
The  reduction of the magnetic field ({\it parameter combination~5}) leads to 
the  increase of   the magnetic radiation length,  as well as 
to the increase  of the magnetic pair production threshold.
As a result,  the cascade develops slower.   
 
Figure~17 illustrates  the  spectral evolution of cascade \grs 
calculated for $H=100 \ \rm G$ and two different temperatures 
of the black-body radiation, $kT=0.3$ and 3 eV 
({\it parameter combinations~3} and {\it 4}). 
It is seen that at small depths ($t\le 10^{10}$~cm for the upper panel 
and $t\le 10^8$~cm for the bottom panel) the   
spectra  are  fully  determined  by interactions with the magnetic
field. At these  depths a ``standard'' power-law 
spectrum with $\alpha=1.5$ is formed  in a broad  energy region 
below $E_{\rm th}^{(F)}$. At larger depths  the interactions 
with  the photon gas start to deform  the shape of the spectrum.
These interactions  lead  to absorption of photons  
above $E_{\rm th}^{\rm (G)}$. At the same time, 
synchrotron radiation  of the photo-produced pairs  
appears  in the energy region $E\le E_{\rm th}^{\rm (G)}$. 
Thus, only the photons with energy 
below $E_{\rm th}^{\rm (G)}$ can survive  at large depths.
The   spectrum of these photons is close to power-law with 
photon  index 1.8-1.9, i.e.  steeper than the  
the canonical $\varepsilon^{-1.5}$  cascade photon spectrum formed     
in the  pure photon gas or  pure magnetic field. 
This is explained by the  break of symmetry
between electrons and $\gamma$-rays. While 
at late stages of the cascade development  \grs continue to produce 
high energy  electron-positron pairs,  the synchrotron cooling of these 
electrons does not anymore support the cascade,  but rather 
destroys it. 

\begin{figure}[htbp]
\centering
\includegraphics[height=0.3\textheight]{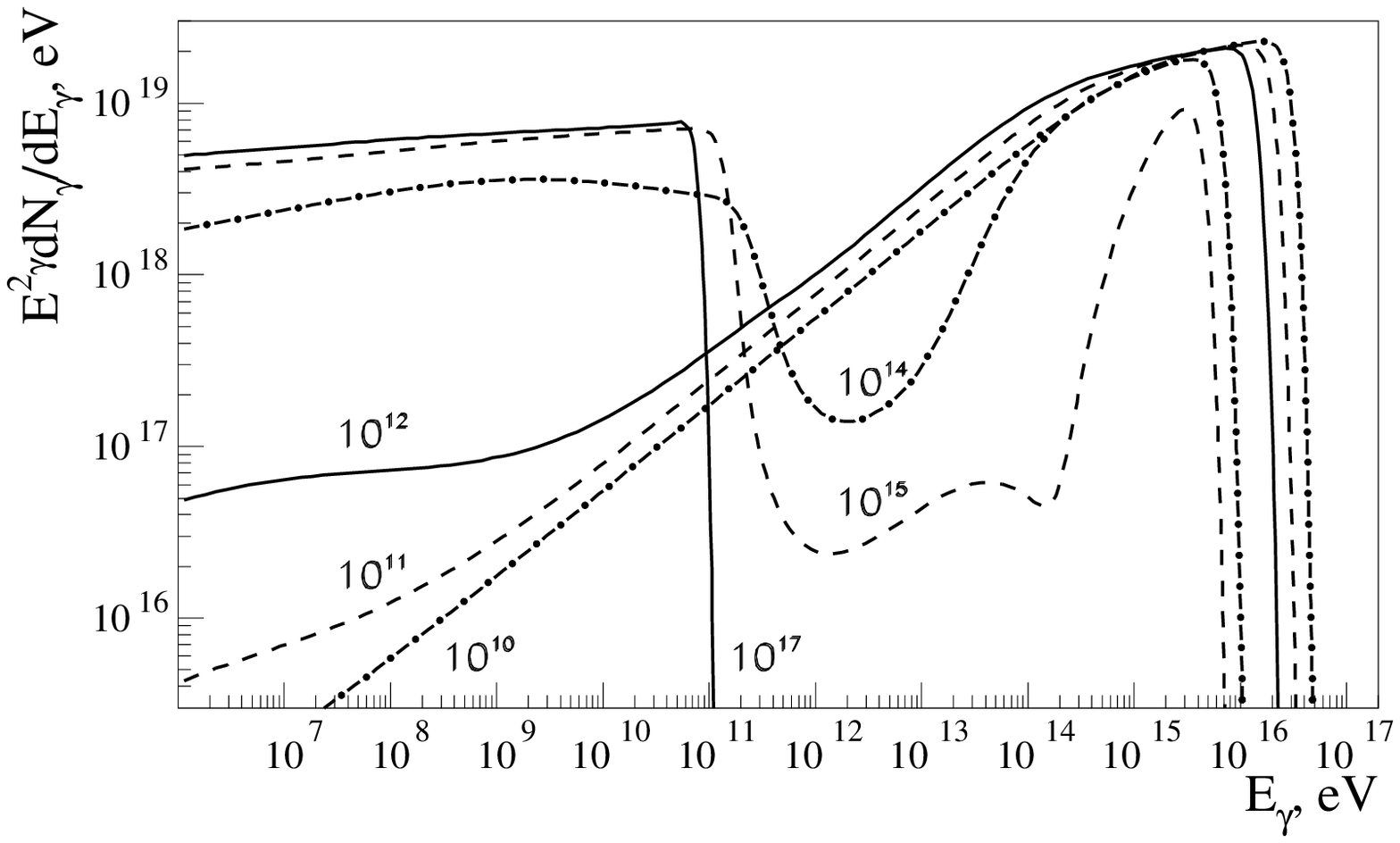} 
\includegraphics[height=0.3\textheight]{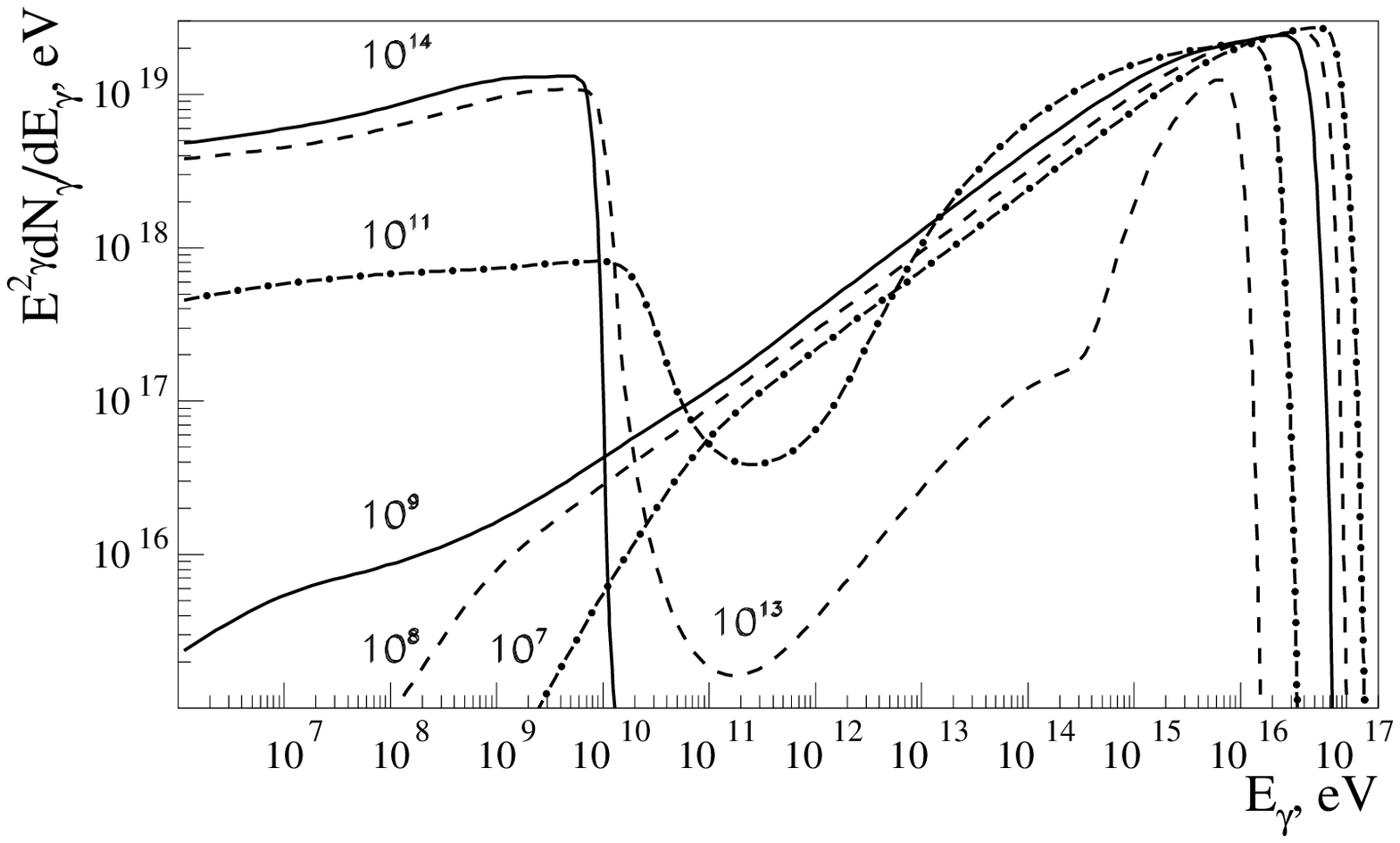} 
\caption{Differential energy spectra of \grs  
of  the  cascade initiated   
by a primary photon  of energy $E_0=10^{20}$~eV
in the   compound environment consisting of a blackbody 
radiation  of temperature $T$ and a homogeneous magnetic field
H=100~G. Upper panel -- $kT=0.3~eV$; bottom panel -- $kT=3$~eV.
The spectra are calculated at different depths (in cm) indicated at the curves.} 
\label{fgr17}
\end{figure}

\begin{figure}[htbp]
\centering
\includegraphics[height=0.3\textheight]{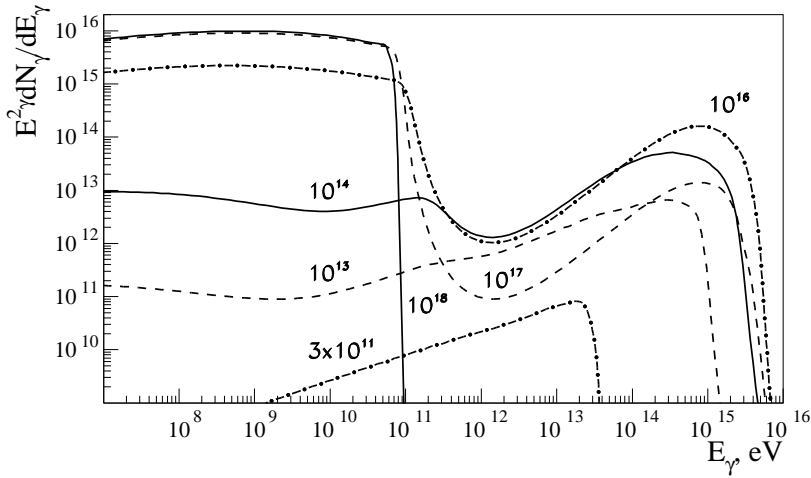}  
\caption{The same as in Fig. \ref {fgr17},
but for the energy of primary photons $E_0=10^{17}$~eV
and the magnetic field $H=1 \ \rm G$. The temperature of the black-body radiation is 
assumed 0.3 eV. 
}
\label{fgr18} 
\end{figure}

Finally, in Fig.~\ref{fgr18} we show  the spectral evolution of 
cascade \grs calculated for a less extreme combination 
of model parameters.  Namely,
compared to Fig.~\ref{fgr17}, we assume   a smaller energy 
for primary \grs ($E=10^{17}$~eV) and weaker magnetic
field ($H=1$~G). At these conditions the primary photons 
do not interact with the magnetic field. Instead,  
the cascade development starts  with  pair production in 
the radiation field. Nevertheless, we can see that many basic features,
in particular the $\varepsilon^{-2}$ type spectrum at low energies, 
are quite similar to the spectra shown in  Fig.~\ref{fgr17}.

% Figure 17.
% ----------------------------------------------------------------------------

% Figure 18.
% ----------------------------------------------------------------------------

\section{Summary}

In this study,   the technique of adjoint cascade 
equations has been applied to  investigate  
properties of electron-photon  cascades 
in  hydrogen gas and  in ambient radiation 
and magnetic fields.
We also have inspected   the main features of 
cross-sections of relevant processes that initiate  and support 
cascade developments  in these substances.

The cascade curves of electrons and photons in the photon gas and magnetic 
field have features quite different  from the 
cascade curves in matter. 
The energy spectra of cascade particles are also considerably different 
from the conventional cascade spectra in matter. The spectra for 
the magnetic field have properties intermediate between those for   
cascade spectra in matter and in the photon field.   
Although  for  certain astrophysical scenarios  the development of cascades in 
``pure'' environments  can be considered as an appropriate  
and fair approximation, at some 
conditions the interference of processes associated with  
interactions of cascade electrons and \grs with the  ambient photon  
gas and magnetic field (or matter) can significantly change the character 
of cascade development, and consequently the spectra of 
observed $\gamma$-rays. The impact is very complex and quite sensitive to 
the choice of specific parameters. Therefore each practical case 
should be subject to  independent studies.

\section*{Acknowledgements}

We are  grateful to A.Timokhin for fruitful discussions. AVP
thanks Max-Planck-Institut f\"ur Kernphysik (Heidelberg) for 
hospitality and support during his work on this paper.


\small

\begin{thebibliography}{100}
%1
\bibitem{matter}
B. Rossi, K. Greisen, Rev. Mod. Phys. 13 (1941) 419; 
J.Nishimura,   {\it Handbuch der Physik} Bd.XLVI/2  (1967) 1;
I.P Ivavenko, {\it  Electromagnetic Cascade Processes} (1968),
Moscow State University  Press (in Russian);
T.K. Gaisser, {\it Cosmic Rays and Particle Physics} (1990), 
Cambridge University  Press.
%2
\bibitem{EGS4}
W.R. Nelson, H. Hirayama,  D. Rogers,  Preprint SLAC-265 (1985),
Standford University.  
%3
\bibitem{Berez_book}
V.S.  Berezinsky,  S.V. Bulanov,  V.A.  Dogiel,  V.L. Ginzburg,  V.S. Ptuskin 
{\it Astrophysics of cosmic rays} (1991), Amsterdam: North-Holland.
%4
\bibitem{Ber_Vest}
V.S.  Berezinsky, in Roberts A. (editor), Proc. 1976 DUMAND Summer Workshop
(1976), FNAL, Batavia, p. 229;    
D. Eichler, W.T. Westrand, Nature 307 (1984) 613. 
%5
\bibitem{Halzen}
F. Halzen, D. Hooper, Rep. Prog. Phys. 65 (2002) 102.
%6
\bibitem{AVK}
F.A. Aharonian, V.V. Vardanian, V.G. Kirillov-Ugryumov,
Astrophysics (tr. Astrofizika) 20 (1984)  118; 
F.A. Aharonian,  V.V. Vardanian, Astr. Sp. Sci. 115 (1985) 31.
%7
\bibitem{Zdz}
A.A. Zdziarski, Astrophys. J. 335 (1988) 786.
%8 
\bibitem{Svenson}
R. Svensson, MNRAS 227 (1987) 403.
%9
\bibitem{CopBl}  
P.S. Coppi,  R.D.  Blandford, MNRAS 245 (1990) 453.
%10
\bibitem{mastiprot}  
A. Mastichiadis, R.J  Protheroe, A.P.  Szabo, MNRAS 266 (1994) 910. 
%11
\bibitem{GRBs}
G. Cavallo, M.J.  Rees, MNRAS  183  (1978) 359;
E.V. Derishev, V.V. Kocharovsky, Vl.V. Kocharovsky, Astron. Astrophys.
372 (2001) 107;
C.D. Dermer, Astrophys. J.  574 (2002)  65.
%12
\bibitem{blazars}
K. Mannheim, Phys. Rev. D 48 (1993)  2408;
R.D. Blandford, A. Levinson, Astrophys. J. 441 (1995), 79;
A.  Atoyan, C.D.  Dermer, Phys. Rev. Lett. 87 (2001) 221102;  
A. Muecke, R.J. Protheroe, R. Engel, J.P. Rachen, T. Stanev, Astropart. Phys. 
(2002), in press;   
%13
\bibitem{bierman}
P.L. Biermann,  P.A. Strittmatter, Astrophys. J. 322 (1987) 643; 
K. Mannheim, P.L. Biermann, W.M. Kruells, Astron. Astrophys. 251 (1991) 723.
%14
\bibitem{gamma_jets}
A. Neronov, D . Semikoz, F. Aharonian, O. Kalashev,  Phys. Rev. Lett. 89 (2001) 1101. 
%15
\bibitem{psynch}
F. A. Aharonian, MNRAS 332 (2002) 215.
%16
\bibitem{halos}
F.A. Aharonian,  P.S.  Coppi, H.J.  V\"olk, Astrophys. J. 423 (1994)  L5.  
%17
\bibitem{intergalactic}
F.A. Aharonian, A.M. Atoyan, Sov. Phys. JETP 62 (1985) 189;
R.J. Protheroe, MNRAS 221 (1986)  769;
F.A. Aharonian, B.L. Kanevsky, V.V. Vardanian, Astr. Sp. Sci 167 (1990) 93; 
I.P. Ivanenko, A.A. Lagutin, Proc  22nd ICRC (Dublin), 1991, vol. 1, p.  121;
F.A. Aharonian,  P.  Bhattacharjee, D.N. Schramm,  Physical Review D 46 (1992) 4188;  
R.J. Protheroe, T.  Stanev, MNRAS (264) 191, 1993; 
R.J. Protheroe, T. Stanev, V.S. Berezinsky, Nucl. Phys. B 43 (1995) 62;
R.J. Protheroe, T. Stanev, Phys. Rev. Letters 77 (1996) 3708;
P.S. Coppi, F.A. Aharonian,  Astrophys. J. 423 (1997)  L9; 
G. Sigl,  S. Lee,  D. Schramm,  P. Coppi, Phys. Letters B 392 (1997)  129;
S. Lee,  Phys. Rev. D  58 (1998) 043004, 
Z. Fodor, S.D.  Katz,  Phys. Rev. D 63 (2001) 023002;
F.A. Aharonian, A.N.  Timokhin, 
A.V.  Plyasheshnikov, Astronon.  Astrophys. 384 (2002) 834;
O. Kalashev, V. Kuzmin, D. Semikoz, G.  Sigl,   Phys. Rev. D 65 (2002) 103003.  
%18
\bibitem{Bo_Rees}
S. Bonometto and M. J. Rees, MNRAS 152 (1971) 21. 
%20
\bibitem{Guilbert}
P.W. Guilbert, A. C. Fabian, M.J. Rees, MNRAS 205 (1983)  593. 
%21
\bibitem{Ah_vard_kir}  
F.A. Aharonian,  V.G. Kririllov-Ugriumov, V. V. Vardanian, 
Astr. Sp. Sci. 115 (1985) 201.
%22
\bibitem{sturrock}
P.A. Sturrock, Astrophys. J. 164 (1971) 529.
%23
\bibitem{HardBaring} 
M.G. Baring, A.K. Harding, Astrophys. J.  547 (2001) 529.
%24
\bibitem{gonchar}
B.L. Kanevsky and A.I. Goncharov, Voprosy atomnoy nauki i techniki 4 (1999) p. 1
(in Russian).
%25
\bibitem{Anguelov}
 V. Anguelov, H. Vankov, J. Phys. G: Nucl. Part Phys. 25 (1999) 1755.
%26
\bibitem{Plyah}
A.V. Plyasheshnikov,  F.A. Aharonian, . J. Phys. G: Nucl. Part Phys. 28 (2002) 267.
%27
\bibitem{Bednarek}
W. Bednarek, MNRAS 285 (1997) 69. 
%28
\bibitem{Akhiezer}
A.I. Akhiezer, N.P. Merenkov, A.P. Rekalo   J.Phys.G: Nucl. Part. Phys.  20  
(1994) 1499.
%29
\bibitem{Landau} 
L. D. Landau, G. Rumer, Proc. R. Soc A 166 (1938) 213. 
%30
\bibitem{Uchaikin}
V.V. Uchaikin, V.V.  Ryzhov  {\it  The Stochastic Theory
of Transport of High Energy  Particles} (1998),  Novosibirsk, Nauka 
(in Russian)                                                                      
%31
\bibitem{Plyasheshnikov}
A.V. Plyasheshnikov, A.A. Lagutin,  V.V.  Uchaikin,  
{\it Proc. of 16th ICRC} (1979), Kyoto, vol.  7, p.1 .
%31
\bibitem{doublpair}
R.W. Brown, W.F. Hunt, K.O. Mikaeilian, I.J. Muzinich,
Phys. Rev. D 8 (1973) 3083.   
%32
\bibitem{triplet}
A. Mastichiadis, MNRAS 253 (1991) 235;
C.D. Dermer, R. Schlickeiser, Astron. Astrophys. 252 (1991) 414. 
%33
\bibitem{adler} 
S.L. Adler, Ann. Phys. 67 (1971)  599. 
%34
\bibitem{Akh_Berest}
A.I. Akhiezer, V.B. Berestetskii, {\em Quantum Electrodynamics}, 
1965, Interscience, New York.
%35
\bibitem{BlumGould}
G.R. Blumenthal, R.G. Gould, Review Mod. Phys. 42 (1971) 237. 
%36
\bibitem{AhAtNah}
F.A. Aharonian, A.M. Atoyan, A.M.  Nagapetyan, 
Astrophysics (tr. Astrofizika) 19 (1983) 187. 
%37
\bibitem{Altai}
A.K. Konopelko, A.V.  Plyasheshnikov,  Nucl. Instr. Meth. 450 (2000) 419.
%
\end{thebibliography}
\end{document}